\documentclass[journal=jacsat,manuscript=article,superscriptaddress]{achemso}
\usepackage[version=3]{mhchem} 
\usepackage[table]{xcolor}
\usepackage{booktabs}
\usepackage{colortbl}
\usepackage[table]{xcolor}
\usepackage{graphicx}
\usepackage{textcomp}
\usepackage{comment}

\author{Chun-Ting Lin}
\altaffiliation{These authors contributed equally.}
\affiliation{Department of Chemistry, The Pennsylvania State University, Pennsylvania, United States of America}

\author{Diganta Dasgupta}
\altaffiliation{These authors contributed equally.}
\affiliation{Condensed Matter and Statistical Physics, The Abdus Salam International Centre for Theoretical Physics (ICTP), Trieste, Italy}
\alsoaffiliation{Scuola Internazionale Superiore di Studi Avanzati (SISSA), Trieste, Italy}

\author{Tinglu Yang}
\affiliation{Department of Chemistry, The Pennsylvania State University, Pennsylvania, United States of America}

\author{Cesare Malosso}
\affiliation{Laboratory of Computational Science and Modelling, IMX, École Polytechnique Fédérale de Lausanne, Lausanne, Switzerland}

\author{Giulia Sormani}
\affiliation{Scuola Internazionale Superiore di Studi Avanzati (SISSA), Trieste, Italy}

\author{Colin Egan}
\affiliation{Initiative for Computational Catalysis, Flatiron Institute, New York, United States of America}

\author{Giovanni Bussi}
\affiliation{Scuola Internazionale Superiore di Studi Avanzati (SISSA), Trieste, Italy}

\author{Ali Hassanali}
\affiliation{Condensed Matter and Statistical Physics, The Abdus Salam International Centre for Theoretical Physics (ICTP), Trieste, Italy}
\email{ahassana@ictp.it}
\phone{+39 040 2240 4175}

\author{Paul S. Cremer}
\affiliation{Department of Chemistry, The Pennsylvania State University, Pennsylvania, United States of America}
\email{psc11@psu.edu}
\phone{(814) 865-6259}

\title[Lithium-Chloride Solutions]
{Beyond the Virial Expansion: Microscopic Origins of Partial Molar Volumes in LiCl Solutions}

\abbreviations{IR,NMR,UV}

\begin{document}






\begin{tocentry}
  \centering
  \includegraphics[width=8.25 cm, height=4.45cm]{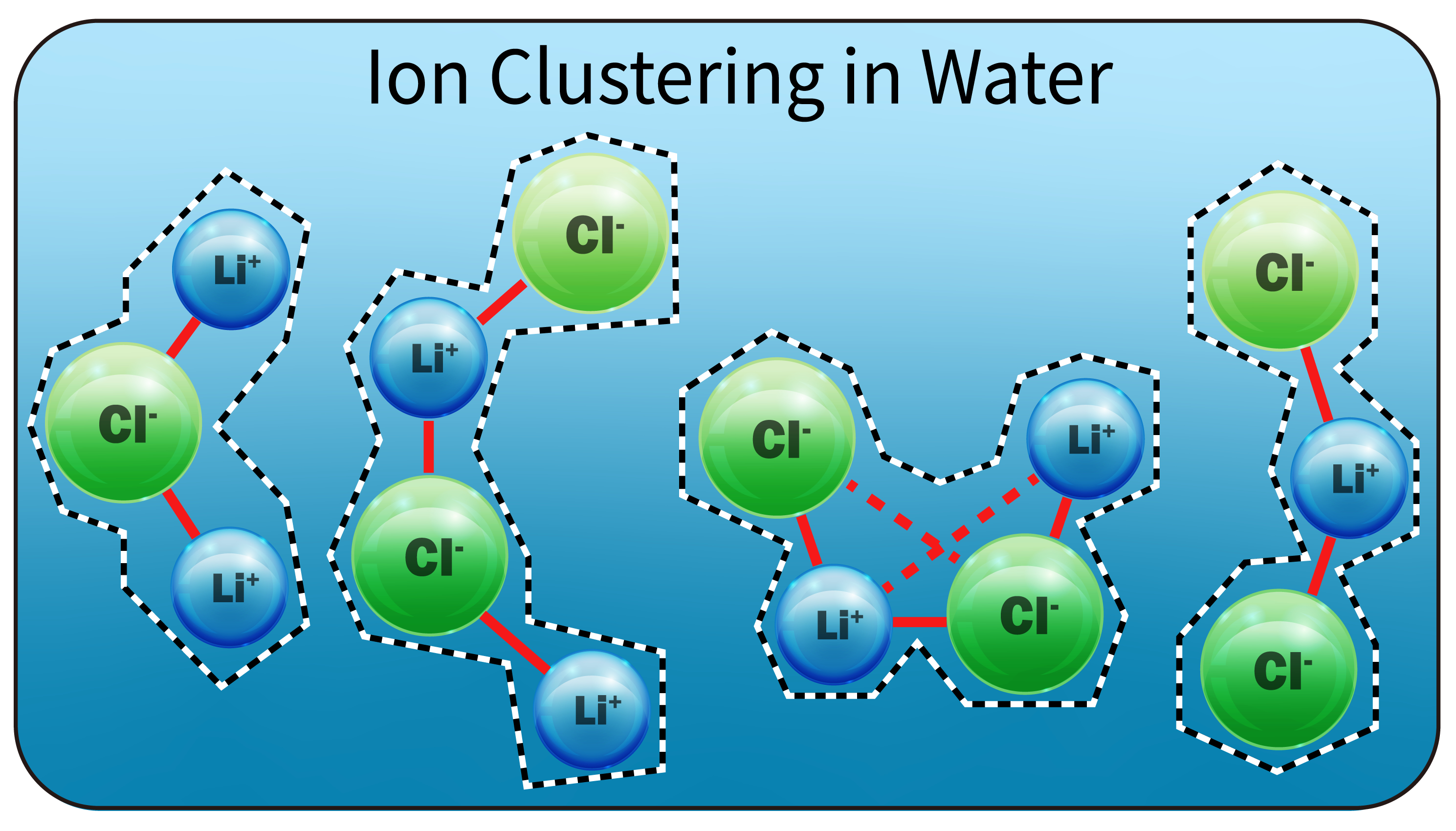}
  \label{For Table of Contents Only}
\end{tocentry}

\begin{abstract}
Although electrolyte density measurements have been reported for over a century, employing them to obtain accurate partial molar volume (PMV) profiles as a function of salt concentration has remained elusive. Obtaining such curves requires precise density measurements combined with a proper treatment of the associated virial expansion. In this work, we obtain PMV profiles for aqueous LiCl solutions. The resulting data enable the development of highly accurate force fields for Li$^+$ and Cl$^-$ ions, revealing a clear progression from isolated ions to ion pairs and ultimately to higher-order chain and ring structures. Because ion clustering emerges from complex, nonlocal interactions, it cannot be easily mapped onto specific virial terms. Instead, a direct structural and volumetric interpretation can be achieved by partitioning molecular dynamic (MD) simulation snapshots into three-dimensional polyhedral regions associated with individual salt ions and water molecules. The corresponding ionic and water volumes from this treatment quantitatively reproduce the experimental PMV curve. The results demonstrate that the PMV for salt increases (while that of water decreases) up to 6.7 M. Above this concentration, the direction reverses as three- and four-body interactions become prominent. Complementary multivariate curve resolution (MCR) Raman spectroscopy and density functional theory (DFT) calculations elucidate the molecular-level details of water electrostriction, which also persists up to 6.7 M. Significantly, the PMV data can be correlated with key thermodynamic properties, including the osmotic coefficient and the eutectic point. The procedures established here provide a general framework for modeling electrolyte solutions and enable the development of a new generation of accurate force fields for aqueous ions.
\end{abstract}


\section{Introduction}

Ion-specific effects in aqueous electrolyte solutions have been actively explored since the foundational studies of Franz Hofmeister\cite{hof} in the 1880s. Their influence is pervasive across the physical, chemical, biological, and material sciences\cite{intro_review_marcus, intro_review_vandervegt, intro_voigt,intro_zhang_review,fayer2022}. For example, the specific identity of both cations and anions influences transport properties in batteries and fuel cells\cite{intro_battery_bian, intro_battery_suo}, affects corrosion rates\cite{intro_corrosion_jiang, intro_corrosion_ma}, solvent extraction efficiency\cite{intro_solvent_extract}, as well as protein and drug molecule crystallization\cite{intro_kunz, intro_protein_sol_kim, intro_zhang_review}. Ion-specific interactions are also at the heart of technologies aimed at optimizing wastewater treatment\cite{intro_wastewater_nemes} and carbon capture\cite{intro_carboncapture_hegarty}. Moreover, cation specificity is central to the behavior of living organisms. For example, K$^+$ and Mg$^{2+}$ accumulate in the cytosol, while Na$^+$ and Ca$^{2+}$ are highly abundant in the extracellular matrix, a crucial factor for controlling the cell membrane potential\cite{intro_lehninger}. While Li$^+$ is less frequently encountered in biological systems, its clinical utility in treating bipolar disorder is well established. Additionally, endogenous trace Li$^+$ levels within cerebrospinal fluid are vital to maintaining cognitive function\cite{intro_li_su, intro_li_alzheimer}. 

To understand electrolyte properties in aqueous solutions, it is necessary to anchor the behavior of dissolved salts to fundamental solution properties, such as density. Indeed, the density of an aqueous solution containing ions has historically been used as a key property to parameterize and validate interaction potentials in all-atom molecular dynamics simulations\cite{density_target_alkane, density_target_booth, density_target_gaff, density_target_madrid_nacl, density_target_opls, density_target_opls_aa, density_target_parm94, density_target_roux, desnity_target_pavel, ion_madrid_2019, water_jorg_tip3_4p, water_spce, water_tip4p_05_abascal}. Curiously however, the vast majority of the experimental density data gathered throughout the 20$^{\text{th}}$ century are neither sufficiently accurate nor precise to obtain reliable partial molar volume (PMV) curves as a function of salt concentration. The challenge is that the PMV for salt ($V_m^{salt}$) and water ($V_m^{H_2O}$) need to be obtained from the derivative of the density with respect to salt concentration. This requires high-quality measurements, which are difficult to make with classical weight and volume measurements, as well as only moderately pure chemicals.

Beyond the challenges of obtaining accurate experimental data, interpreting thermodynamic behavior introduces an additional layer of complexity. Building on the statistical foundations established by McMillan and Mayer\cite{mac_mayer_1, mac_mayer_2}, the virial expansion has become the standard framework for describing how thermodynamic properties of electrolyte solutions, such as the PMV, depend on concentration (Equation \ref{eqn:virial}):
\begin{equation}\label{eqn:virial}
    V_m = B_0 + B_1[c] + B_2[c]^2 + B_3[c]^3 + B_4[c]^4
\end{equation}

In this equation, B$_0$ represents the partial molar volume at infinite salt dilution, while the higher order terms, B$_n$, with $n>0$, serve as correction factors that describe how the PMV evolves with increasing salt concentration. From a statistical mechanical perspective, the virial coefficients can be formally related to contributions from one-body, two-body, up to n-body correlations\cite{mac_mayer_1, intro_hansen_mcdonald}. The main conceptual difficulty lies in their microscopic interpretation. Although the PMV is a well-defined thermodynamic property, mapping the B$_n$ coefficients onto physically meaningful local volumes associated with specific configurations of ions and water molecules is not straightforward. This challenge becomes particularly pronounced at higher electrolyte concentrations, where many-body correlations, collective behavior, and cooperative effects play an increasingly significant role.

\begin{figure}[ht]
    \centering
    \includegraphics[width=0.75\linewidth]{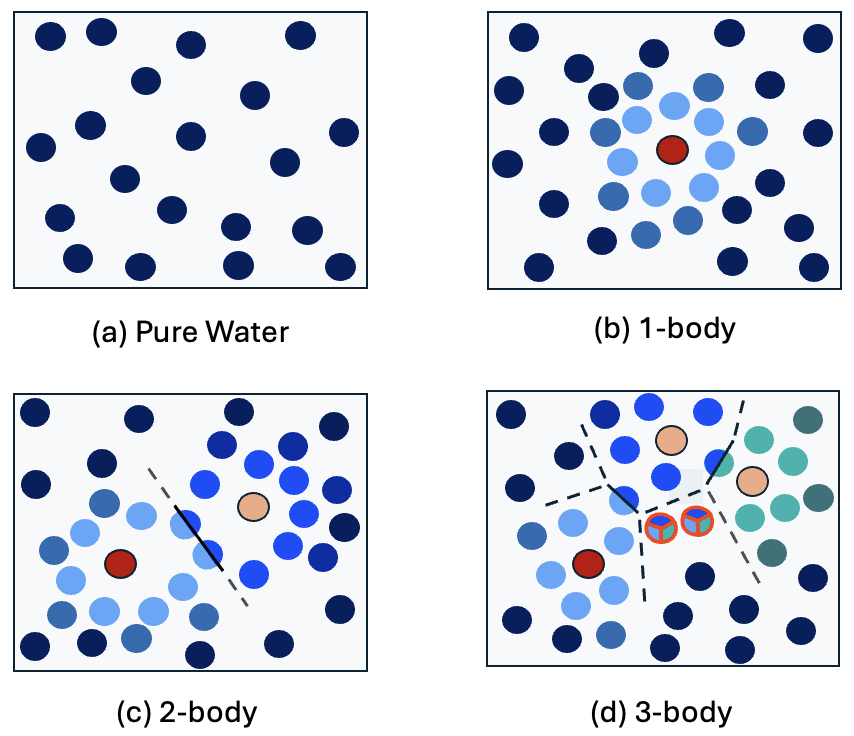}
    \caption{\textit{Schematic diagrams of the interactions between ions and water molecules in aqueous solutions showing a structural progression from (a) water molecules in neat water (navy blue spheres) to (b) the hydration of a single ion (red sphere), and finally to increasingly complex solvent structuring as (c) one and (d) two counterions (beige spheres) are introduced. Water molecules in the hydration shells of the ions are shaded light blue, bright blue, and light green. When a water molecule is in the hydration shell of more than one ion, multiple colors are used. }}
    \label{fig:schematic}
\end{figure}

Figure \ref{fig:schematic} illustrates the evolution of an electrolyte solution upon the introduction of salt, starting from pure water. A variety of multi-body effects directly include the response of the surrounding solvent molecules as well as the ions. The question arises as to how such n-body interactions contribute to the PMV. Moreover, one can ask how the structure of a 2-body or 3-body effect is defined at the molecular level. These questions currently remain open in electrolyte theory but have profound implications for a variety of thermodynamic properties such as viscosity, conductivity, surface-tension and osmotic pressure, among others, that can each be fit to a power series. Molecular dynamics simulations could, in principle, provide detailed structural insight to clarify the interpretation of these parameters, but this requires interatomic potentials capable of accurately reproducing experimental PMVs, a requirement that has remained challenging to meet.
\hspace{0.3cm}
\hspace{0.3cm}

Herein, a general framework is presented for probing electrolyte solutions that combines thermodynamic measurements, vibrational spectroscopy, molecular dynamics simulations and quantum-chemistry calculations to obtain and interpret the PMV for electrolyte solutions. The procedure begins with state-of-the-art pycnometry measurements of the density, enabling the extraction of $V_m^{salt}$ and $V_m^{H_2O}$ as a function of salt concentration. These measurements afford aqueous electrolyte solution densities to five significant figures at ±0.03\% uncertainty. Next, we conducted all-atom classical molecular dynamics (MD) simulations using the TIP4P/2005\cite{water_tip4p_05_abascal} water model. This was done via an optimization procedure in which the ion force-field parameters were tuned to reproduce the experimentally measured solution density and PMV values as a function of concentration. The data harvested from the MD simulations were used to build a PMV model, bottom-up, from local Voronoi-like volumes\cite{voro_paper, voro_laguerre_main, tess_1} associated with ions and water. Finally, the response of the hydrogen-bonding network of water was probed as a function of salt concentration using Raman spectroscopy analyzed via multivariate curve resolution (MCR)\cite{raman_mcr_review, raman_mcr_1_dor, raman_mcr_2_dor}. The OH stretch resonances were assigned by normal-mode vibrational spectra determined from density functional theory (DFT) electronic structure calculations\cite{raman_theory_hess}.

This synergistic experimental and computational approach was used to investigate LiCl solutions between 0.1 M to 9.5 M at room temperature and pressure. Due to its small ionic radius and high charge density, Li$^+$ served as a versatile model to investigate how kosmotropic cations\cite{intro_zhang_review, intro_review_marcus} perturbed the thermodynamic and dynamical properties of salt solutions. It was found that the PMV curves for salt and water displayed a maximum and minimum, respectively, near 6.7 M. Using molecular simulations from optimized potentials that nearly quantitatively reproduced the experimentally determined PMVs, we demonstrated that non-local ion-pairing led to the formation of chain and ring structures that were a crucial factor shaping the PMV data. By decomposing the thermodynamic parameters into molecular contributions, these results indicated that the PMV arose from a collective response involving both isolated hydrated ions and correlated species, forming non-local ion pairs. Vibrational Raman spectra revealed distinct populations of waters engaged in weaker and stronger hydrogen bonds, which DFT calculations link to distorted hydrogen-bond geometries and highly polarized water molecules confined between ion pairs, respectively.


\section{Methods}

We begin by briefly summarizing the essential aspects of the experimental and computational methods that were employed in this work before proceeding to the results. A more detailed description of the equipment, protocols and procedures which were used on both the experimental and theoretical fronts can be found in the Supporting Information (SI).

\subsection{A. Density Measurements}

Aqueous LiCl solution density was measured at 20.0$^\circ$C using a borosilicate glass pycnometer with a calibrated volume of 53.372 $\pm$ 0.002 mL. A NANO pure water system (Barnstead) and a Milli-Q-UF-Plus water purifier (Millipore) were used to obtain 18.2 M$\Omega$·cm water, which was, in turn, employed to prepare LiCl solutions. To begin an experiment, LiCl (ACS reagent, $>$ 99\%) was baked at 580$^\circ$C for 14 hours. The baked salt was then used to prepare electrolyte solutions. Impurities and particulates were removed from the nascently formed solutions by two rounds of syringe filtration. The combined uncertainty in the overall procedure for making density measurements from 0.1009 M to 9.5184 M was less than $\pm$ 0.0005 g/mL. The partial molar volumes were subsequently derived from the density profiles as described in the Results section. For additional details on the density measurements and extraction of the PMVs see Section S1 in the SI.

\subsection{B. Raman MCR Analysis}

A home-built instrument was used to measure Raman spectra. Changes in the water structure in the OH stretch regime were observed between 3000 cm$^{-1}$ to 3800 cm$^{-1}$ as a function of LiCl concentration. MCR analysis was employed to deconvolute individual spectra into a linear combination of solute-correlated ($C_S S_S^T$) and bulk water ($C_W S_W^T$) contributions as well as residual noise (\textbf{E}), starting from an input Raman spectrum matrix (\textbf{D}). The $C_i$ and $S_i^T$ terms denote the concentration profile and resolved component spectra of the Raman signal, respectively, for the solute-correlated and bulk-like water components of the spectra. Further details on the Raman measurements and analysis of the various modes are in Section S2 in the SI.

\subsection{C. Computational Methods}

Classical molecular dynamics (MD) simulations with empirical potentials were used to study the behavior of LiCl solutions as a function of concentration. Specifically, we focused on optimizing the Lennard-Jones (LJ) parameters for Li$^+$ and Cl$^-$ starting from the Joung-Cheatham\cite{ion_jc_2008} ion forcefield. This builds on earlier studies that combine ensemble-based methods with force field refinement techniques\cite{optim_hummer_1, optim_hummer_2, optim_hummer_3, optim_bussi_1, optim_bussi_3, optim_li_1}. TIP4P/2005\cite{water_tip4p_05_abascal, water_tip4p_transfer_dopke} was chosen as the water model and left unmodified. The optimized ion parameters reproduced the experimental densities and PMVs. Next, microsecond long MD simulations over a range of concentrations from 0.5 to 10 M were conducted. The GROMACS 2023.3\cite{software_gmx} package was used for all MD simulations. Temperature and pressure were maintained via the Stochastic Velocity\cite{bussi_thermo} and Cell\cite{bussi_baro} Rescaling recipes, respectively. The results were harvested to perform structural analyses using a combination of Python scripts. The libraries used are referenced herein.\cite{python_numpy, python_scipy, python_net_x, python_pyvoro, software_voro++}.  Raman calculations were conducted on model ion-water clusters using the ORCA\cite{software_orca} package with DFT\cite{orca_d3_grimme, orca_d4, orca_hessian, orca_libxc, orca_ri, orca_shark, orca_vdw, quantum_r2scan, basis_1, basis_2, basis_3, raman_theory_scaling}. All representations of molecular geometries were generated using Visual Molecular Dynamics\cite{software_vmd_main} (VMD). Unless stated otherwise, details of the computational methodologies employed, including optimization protocols and simulation parameters, are described in Section S3 in the SI.


\section{Results}

\subsection{A. Density and Partial Molar Volumes}

\begin{figure}
    \centering
    \includegraphics[width=\linewidth]{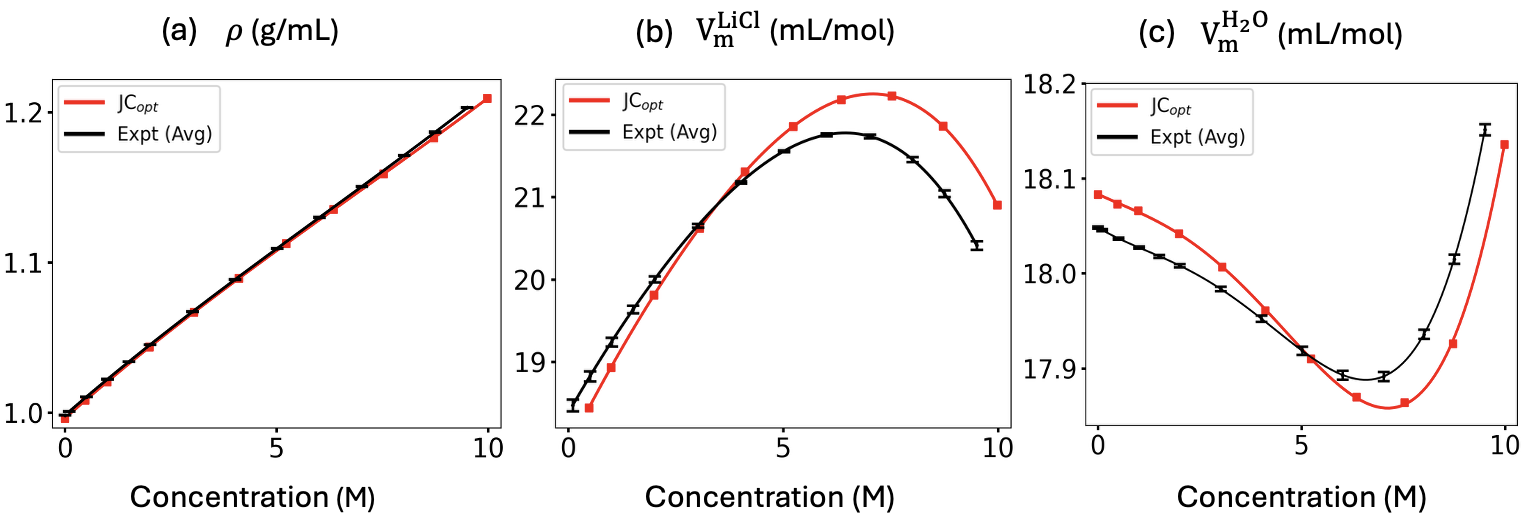}
    \caption{\textit{Experimental (black lines) and simulation (red lines) profiles for (a) density as well as the partial molar volumes of (b) LiCl and (c) water. The experimental data points are drawn with error bars that come from the standard deviation from three independent trials. Uncertainties in the theoretical estimates are smaller than the data points.}}
    \label{fig:dens}
\end{figure}

The experimental density profile for LiCl in water from 0 to 9.5 M is shown in Figure \ref{fig:dens}a (black data points). In the absence of salt, the density of pure water at 1 bar and 20 °C was 0.99818 $\pm$ 0.00008 g/mL. The density increased roughly linearly at low LiCl concentrations but began to bend downward at higher salt concentrations. Fitting the data required a virial expansion up to third order in the density (see Figure S1.1 and accompanying discussion in the SI). The Lennard-Jones parameters for LiCl in water system were then optimized to reproduce the experimental results (red data points and curve, see SI Section S3.1 for details). Next, $V_m^{LiCl}$ and $V_m^{H_2O}$ were derived from the experimental and simulation data in Figure \ref{fig:dens}a by using Equations \ref{eq:form_vm_salt} and \ref{eq:form_vm_wat}, respectively :
\begin{subequations}
\begin{align}
V_m^{LiCl} &= \frac{M_{LiCl}}{\rho}-\frac{1}{\rho^2}\left(1000 + mM_{LiCl}\right)\left(\frac{\partial\rho}{\partial m}\right) \label{eq:form_vm_salt} \\ 
V_m^{H_2O} &= \frac{M_{H_2O}}{\rho}+\left(\frac{mM_{H_2O}}{\rho^2}\right)\left(1 + \frac{mM_{LiCl}}{1000}\right)\left(\frac{\partial\rho}{\partial m}\right) \label{eq:form_vm_wat}
\end{align}
\end{subequations}

Here, the $M_i$ parameters represent the molar mass of LiCl and water, respectively, while \textit{m} is the molality of the solution (see Section S1.2.A in the SI for the derivation of the PMV). The term, $\partial \rho / \partial m$, is the derivative of the solution density with respect to molality. Note that while molality must be employed to generate the PMV curves, the data are converted to units of molarity in Figure \ref{fig:dens} to align with the constant-volume basis of density measurements. The thermodynamic PMV employed here differs fundamentally from the apparent molar volume commonly reported in the literature, especially at high salt concentrations (see Sections S1.2.A and S1.3A in the SI for additional discussion on this point).

\hspace{0.3cm} 
\hspace{0.3cm}

Changes in $V_m^{LiCl}$ and $V_m^{H_2O}$ as a function of the LiCl concentration are provided in Figures \ref{fig:dens}b and \ref{fig:dens}c, respectively. The black points and curves are from experimental measurements. Significantly, $V_m^{LiCl}$ rose continuously from low concentration to $\sim$6.7 M, where it reached a maximum. Above this value, it began to decrease. $V_m^{H_2O}$ showed the opposite trend with a minimum at $\sim$6.7 M. Such reciprocal behavior is a direct consequence of the Gibbs–Duhem relation\cite{gibbs_duhem_Gokcen}, which thermodynamically couples the PMV values for multi-component systems. The pronounced nonlinearity observed in both PMV curves underscores the significance of higher-order interactions in concentrated solutions. Indeed, fitting the PMV data required terms up to 4$^{\text{th}}$ order from Equation \ref{eqn:virial} for both the salt and water to accurately model the data over the entire concentration range (see Section S1.2.B in the SI for details). 

A microscopic interpretation of the PMV trends requires theoretical insight from MD simulations employing force fields capable of reproducing the trends. In density-driven parametrization, it is standard practice to refine the Lennard–Jones (LJ) parameters, $\epsilon$ and $\sigma$, associated with the components (here Li$^+$ and Cl$^-$) to match experimental data. This could be done equally well for the density and the PMV by using interaction potentials having either scaled\cite{ion_madrid_2019} or unscaled charges\cite{ion_jc_2008} (see Section S3.4 in the SI). In the ensuing analysis, unscaled charges were selected. The resulting trends for solution density and the PMV with the optimized FF (\textbf{JC}$_{\text{opt}}$) are shown as red data points and curves in Figure \ref{fig:dens}. The level of agreement between the experiment and simulation is striking and cannot be achieved with off-the-shelf force-fields (see Section S3.4.A in the SI). Moreover, \textbf{JC}$_{\text{opt}}$ semi-quantitatively captured both the maximum in the PMV for salt as well as the minimum with water at approximately 7 M, which was only $\sim$4.5\% higher than the experimental values. As described below, these features arose from structural changes involving correlated ions, such as those produced by ion-pairing interactions.

\subsection{B. Evolution of Ion Networks}

One of the most widely used approaches for characterizing liquid structure is the analysis of the local coordination environment. To this end, 1 $\mu$s NPT simulations were employed to determine the radial distribution functions (RDFs - g(r)) between different species, as well as their corresponding radial integrals - the coordination numbers (CN’s)\cite{rdf_allen_tildsley}. For each ion-ion and ion-water pair, CN was evaluated by integrating g(r) up to the first minimum, thereby including the first hydration shell of the ions. 
\hspace{0.3 cm}
\begin{figure}
    \centering
    \includegraphics[width=\linewidth]{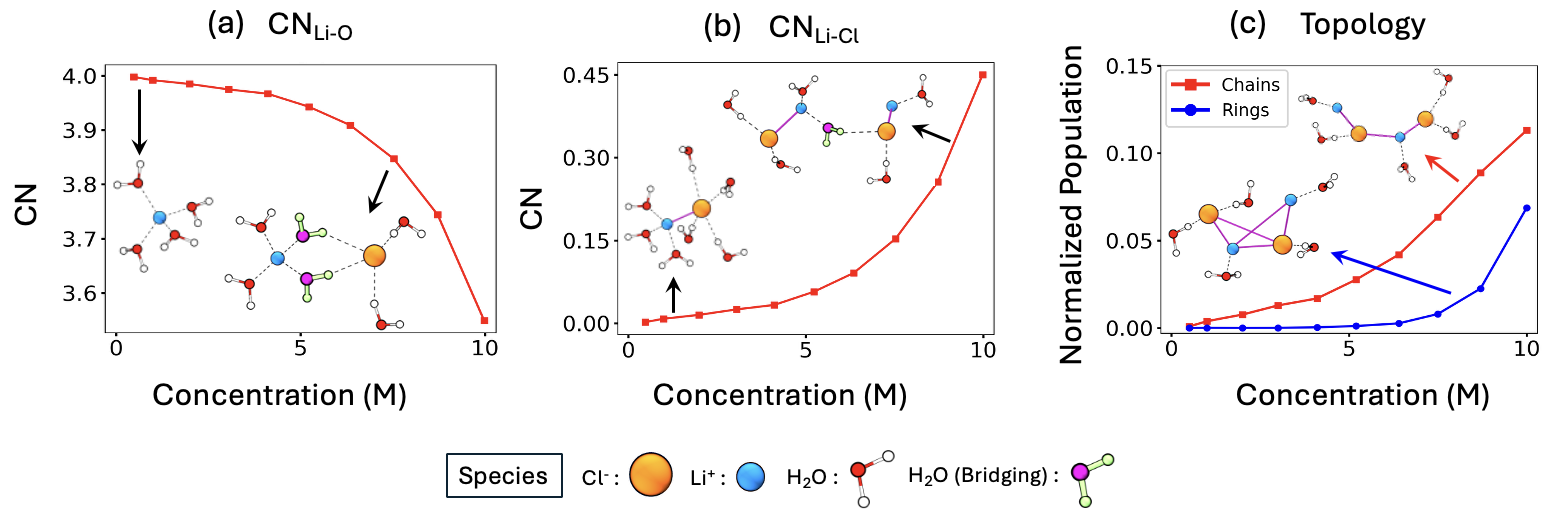}
    \caption{\textit{Ion network reorganization with concentration: (a) Li-O first-shell CN, showing the evolution from isolated, fully solvated ions to ion-pair configurations; (b) Li-Cl first-shell CN, highlighting reduced separations between ion pairs facilitating higher-order aggregate formation; (c) Ring and chain populations in the ion network with representative examples in the insets. The populations are normalized against the number of ions at the corresponding concentration. The cutoff distances for defining these structures provided in Table S3.4}}
    \label{fig:cn}
\end{figure}
\hspace{0.3 cm}

Figures \ref{fig:cn}a and \ref{fig:cn}b illustrate the CNs for Li-O and Li-Cl, respectively, which follow the changes in local structure around the cation with increasing salt concentration. As expected, the Li-O CN falls, while the corresponding value for Li-Cl rises as water is displaced around Li$^+$ by Cl$^-$ at higher salt concentration. The structural evolution in CN is illustrated by snapshots that show how a hydrated Li$^+$ coordinated by 4 water molecules evolves into structures consisting of contact ion pairs and solvent-shared ion pairs\cite{intro_review_vandervegt} (CIP and SIP). Significantly, CN$_{\text{Li-O}}$ and CN$_{\text{Li-Cl}}$ both present rapid changes beyond 5 M after which the maximum/minimum in the PMVs are also observed (Figures \ref{fig:dens}b and \ref{fig:dens}c, $\sim$6.7 M). For CN$_{\text{Li-Cl}}$, one begins to see pairs of CIPs sharing water molecules at higher LiCl concentrations (upper-right schematic in Figure \ref{fig:cn}b). This can be more precisely quantified by examining the CNs associated with the Li–Li and Cl–Cl g(r)’s (see Figures S3.5 and S3.6 in Section S3.4.B in the SI). Beyond the inflection point at $\sim$7 M, a strong tendency is observed for like-ions to come into close proximity.

A more quantitative framework for analyzing structural changes requires moving beyond simple pair-wise correlations. A useful strategy is to map atomic configurations into a network of connected nodes and then apply graph-theoretical analysis\cite{graph_choi_1, graph_choi_2, graph_choi_3, graph_roux}. To this end, we built chemical graphs, treating ions (Li$^+$ and Cl$^-$) as nodes and introducing edges between them based on their physical proximity, determined from the RDFs (see Section S3.5.A in the SI). The advantage of a graph-based representation is that it allows non-local connectivity patterns to be identified. In contrast to RDFs or CNs, graphs inherently capture non-local associations beyond simple pairwise interactions, making it possible to directly visualize the emergence of extended clusters and networks within a defined cutoff distance. Once connectivity is defined, the configuration is naturally partitioned into ion clusters and aggregates of various sizes. For instance, an ion pair consisting of Li$^+$ and Cl$^-$ in direct contact corresponds to a cluster of size i = 2, while larger aggregates reflect an increasing extent of ion aggregation, forming motifs that involve multiple interconnected ion pairs (see Section S3.5 in the SI for further discussion).

Two of the most common topological structures examined in graph-theory are the formation of closed rings\cite{graph_ring_goet, graph_ring_prim} (primitive) and open connected chains\cite{graph_paths_main, graph_paths_1} (paths/walks) . The propensity to form these structures as a function of concentration is plotted in Figure \ref{fig:cn}c. Chain and ring motifs are rare below 6 M; however, above this threshold their concentrations rise sharply. The schematics in Figure \ref{fig:cn}c illustrate two trimer-like rings and a chain, respectively, consisting of two ion pairs. In both cases, the creation of these topological motifs is facilitated by attractive forces between oppositely charged ions that offset local electrostatic repulsions between like-charged ions.

In addition to elucidating the topology of the evolving ion networks, the systematic classification of ions into clusters/aggregates of different sizes provides a rigorous microscopic basis for interpreting experimentally measured PMVs, thereby linking molecular-level structure and organization with macroscopic thermodynamic properties. This idea is explored in detail in the next section.

\subsection{C. Microscopic Origins of the PMV}

\begin{figure}
    \centering
    \includegraphics[width=\linewidth]{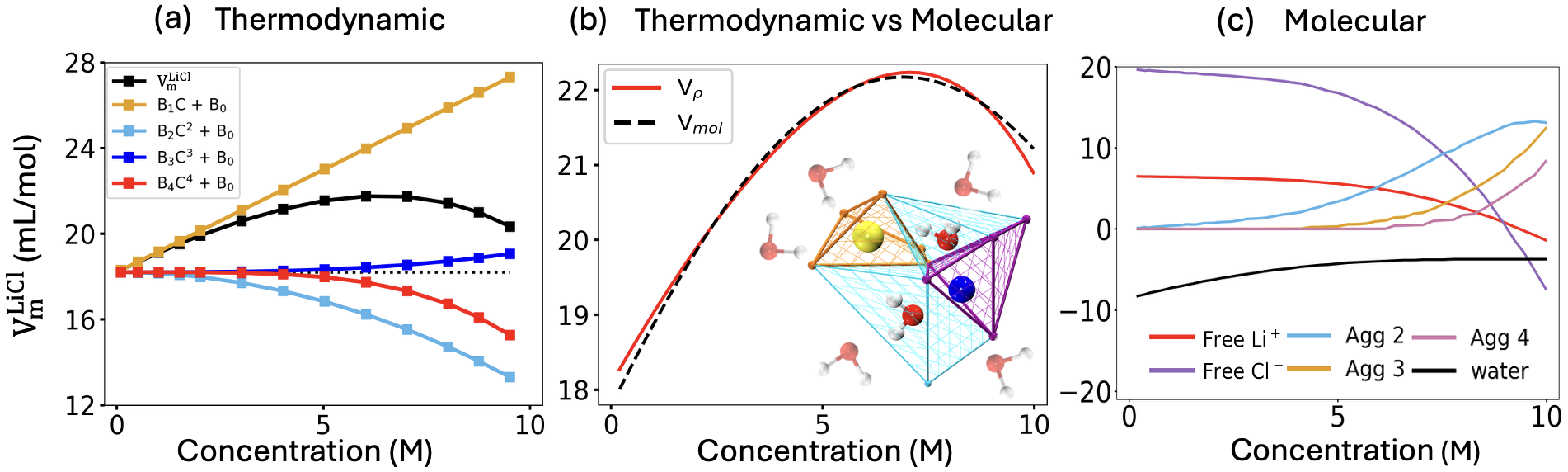}
    \caption{\textit{(a) Evolution of the thermodynamic virial expansion components for $V_m^{LiCl}$ with concentration starting from B$_0$; (b) Comparison between $V_m^{LiCl}$ curves from density derivatives and the Laguerre polyhedral volumes (inset : example of said polyhedron); (c) $V_m^{LiCl}$ contributions from aggregates of different sizes (solvated ions up to a size of 4): ion cluster trends from the bracketed terms in Equation 4, water from the final term.}}
    \label{fig:vmol}
\end{figure}

The PMV virial expansion shown in Equation \ref{eqn:virial} is typically interpreted as a perturbative correction to an ideal reference system. Figure \ref{fig:vmol}a decomposes the experimental salt PMV (black data points and curve) into separate contributions from an initial value at infinite dilution (B$_0$) plus the linear (B$_1$), quadratic (B$_2$), cubic (B$_3$) and quartic (B$_4$) virial terms. In contrast to the case for density, where the intercept term of the virial expansion would represent the density of pure water (independent of ion population), the B$_0$ term for the PMV represents a first-order (one-body) ion contribution. In other words, it reflects the volume change associated with introducing an infinitesimal amount of LiCl into pure water - capturing both the intrinsic Van der Waals volume of the ions and the net electrostriction of the surrounding water molecules. In fact, B$_0$ is directly related to the ratio of the zero order, A$_0$, and first order, A$_1$, virial coefficients of the density (see Section S1.2.C in the SI for details) :  
\begin{equation}\label{eqn:coeff_rel}
    B_0 = \frac{1}{A_0}\left\{M_{LiCl} - 1000\left(\frac{A_1}{A_0}\right)\right\}
\end{equation}

The higher-order coefficients (B$_1$-B$_4$) have alternating signs as can be seen in Table 1 and Figure \ref{fig:vmol}a. This reflects a competition between expansion and contraction effects as the concentration of ions increases. Specifically, B$_1$ (dark yellow points and line, a 2-body effect) has a positive sign representing an increase in the measured ionic volume. The expansion arises from a competition for individual water molecules between two ions as depicted by the multicolored water spheres (dark blue/navy blue spheres) in Figure \ref{fig:schematic}c. In a viral expansion, this term increases the volume continuously with concentration, so that a negative quadratic term (Figure \ref{fig:vmol}a, B$_2$, light blue points and curve reflecting 3-body effects) is required to offset $V_m^{LiCl}$ back to the measured result (black data points and curve). As can be seen, B$_2$ becomes significant near 2.5 M.  Next, the positive B$_3$ and negative B$_4$ contributions (dark blue and red points and curves in Figure \ref{fig:vmol}a) are much less pronounced before $\sim$6.7 M. These trends underscore the fact that higher-order contributions (beyond 3-body effects) only become significant at substantial salt concentrations. Curiously, although the value of B$_3$ is larger than B$_4$ (Table \ref{tab:fit_coefficients}), the PMV contribution for the latter dominates due to the difference in cubic and quartic concentration dependence.  

\begin{table}[h!]
\centering
\caption{Average Fit Coefficients}
\label{tab:fit_coefficients}
\begin{tabular}{|c|c|c|c|c|c|}
\hline
\rowcolor{blue!30} 
Coefficient & B$_0$ & B$_1$ & B$_2$ & B$_3$ & B$_4$ \\
\hline
Value & 18.3815 & 0.9077 & -0.0506 & 0.0009 & -0.0003 \\
\hline
\end{tabular}
\end{table}

These coefficients cannot directly elucidate the microscopic origins of the PMV. Specifically, they fail to succinctly capture how molecular effects such as ion-pairing and the evolution of the ion-water network manifest themselves at different length scales. Schematics like those shown in Figure \ref{fig:schematic} depicting non-local effects involving both ion and solvent reorganization do not provide a quantitative probe for how the PMV evolves at the molecular level. Furthermore, changes in the volume of water as a function of concentration are only implicitly accounted for through the B$_n$ coefficients in the virial expansion.

\hspace{0.3 cm}
\hspace{0.3 cm}

To develop a molecular level model for the behavior of the PMV as a function of salt concentration, it is necessary to quantify the molecular volumes associated with individual water molecules and ions in solution. This can be done by analyzing configurations produced by MD simulations with spatial decomposition methods, such as Voronoi tessellation. This technique seamlessly partitions space into non-overlapping polyhedral cells associated with individual particles and has been widely used to define local geometric volumes in molecular systems\cite{tess_1, tess_2, tess_3, tess_5, tess_6, tess_7}. Here, we employed a weighted variant - radical Voronoi (or Laguerre) tessellation\cite{voro_laguerre_main, tess_1, tess_5},  - to account for atomic radii as weights. This procedure partitions the total volume into individual sub-regions which represent local volumes for ions and water molecules, assigning larger volumes to species with larger radii and vise versa (see Section S3.6 in the SI).

Using Voronoi tessellation, it is possible to reconstruct the PMV from isolated ions, water molecules, and ion clusters that were inherited from the construction of the chemical graphs introduced in the last section. Dissecting the system into contributions arising from ion clusters of different sizes provides a chemically intuitive means to rationalize the PMV in terms of multiple ion-pairs. In contrast, the definitions of clusters in the virial expansion\cite{mac_mayer_1} incorporate both direct and indirect correlations and are therefore not limited to groups of particles within a geometric contact distance, which make them challenging to interpret.

The total volume can be written as a sum of the microscopic volume contributions of the individual components in the mixture: ions (Li$^+$ and Cl$^-$) plus water molecules. As such, it can be shown (see Section S3.6.A in the SI) that the PMV for salt is equivalently given by the following expression (Equation \ref{eqn:vmol}) :
\begin{equation}\label{eqn:vmol}
V_{m}^{LiCl} = 
\underbrace{\sum_{i\in[Agg]} \left[f_i(m) \langle V_{\mathrm{mol}}(m) \rangle_i + m \frac{\partial}{\partial m}\left\{f_i(m) \langle V_{\mathrm{mol}}(m) \rangle_i \right\}\right]}_{\text{Ion Contribution}}
+ 
\underbrace{\frac{1}{M_{H_2O}} \frac{\partial \langle V_{\mathrm{mol}}(m) \rangle_{H_2O}}{\partial m}}_{\text{Water Contribution}}
\end{equation}

\hspace{-0.62 cm}where m is the solution molality, $M_{H_2O}$ is the molar mass of water (in kg), $f_i$ denotes the population fraction of species i, normalized by the total count of LiCl formula units and $\langle V_{mol}\rangle_i$ represents the corresponding average microscopic volumes. The index i spans all aggregate types present at a given concentration starting with isolated (solvated) Li$^+$ and Cl$^-$ that dominate at low concentration. Ion contributions to the PMV, partitioned into different cluster sizes, are controlled by the product of $f_i$ and $\langle V_{mol}\rangle_i$. While $f_i$ values for single ions drop as a function of concentration, the contributions from dimers, trimers and tetramer all grow appreciably beyond 5 M as shown in Figure S3.11 in the SI.

The microscopic volume terms,$\langle V_{mol}\rangle_i$, vary only modestly as a function of salt concentration (see Figure S3.13 in the SI), while changes in $f_i$ are often more substantial. Equipped with both $f_i$ and $\langle V_{mol}\rangle_i$, we can construct the PMV. Figure \ref{fig:vmol}b compares the PMV trends obtained from simulations of the macroscopic density (solid red line), with those derived from the Voronoi-like molecular volumes (black dashed line). The PMV built from clusters ranging from size 1 (isolated ions) to 4 (rings and chains) closely reproduces the thermodynamic PMV. Although models restricted to smaller aggregates qualitatively capture the concentration dependence of the PMV, inclusion of clusters up to size 4 yields a more physically consistent and quantitatively accurate description of the thermodynamic PMV (Section S3.6.C in the SI).

Having established a microscopic model of molecular volumes that quantitatively match the thermodynamic PMV, we can now analyze how clusters of varying sizes contribute to the total volume as a function of concentration. Figure \ref{fig:vmol}c breaks down the PMV by cluster size (1-4) as well as the total contribution arising from the water. Strikingly, the net PMV involves compensating contributions arising from the various-sized clusters, all of which change their volume contribution as a function of salt concentration. For example, the hydrated free Cl$^-$ contribution decreases markedly from 19.6 mL/mol at infinite dilution to 13.0 mL/mol at 6.7 M, while Li$^+$ only shows a modest reduction from 6.5 to 4.4 mL/mol (Figure \ref{fig:vmol}c, purple and red curves). The two-body contribution, which corresponds to single ion-pairs, grows continuously above 0.1 M, increasing to a value of $\sim$7.0 mL/mol close to 6.7 M (Figure \ref{fig:vmol}c, blue curve). Thus, the decrease in hydrated ion contributions is almost exactly offset by the increase of the two-body contribution below 6.7 M. These features are also fully consistent with the structural adjustments associated with ion solvation and ion-pairing (see Figure \ref{fig:cn} a/b). 

For the water contribution in Equation \ref{eqn:vmol}, all water molecules in the simulation box are treated as a single collective entity to circumvent combinatorial complexity. The simulation results show that the $\sim$4.4 mL/mol increase in $V_m^{LiCl}$ from 0 to 6.7 M (Figure \ref{fig:vmol}b) originates from competitive electrostriction of water molecules (Figure \ref{fig:vmol}c, black curve). Indeed, electrostriction weakens with increasing salt concentration, and its attenuation is primarily responsible for the rise in $V_m^{LiCl}$ up to the maximum. Beyond 6.7 M, the water contribution saturates and remains constant. This suggests that electrostriction is complete at this point. In the high salt concentration regime, water molecules are shared among multiple ions, and further addition of salt promotes ion pairing rather than solvation.


Above the maximum, $V_m^{LiCl}$ decreases by approximately 1.4 mL/mol between 6.7 and 10.0 M (Figure \ref{fig:vmol}b). Over this concentration range, the contributions of individual Cl$^-$ and Li$^+$ ions decrease substantially, by 20.4 and 5.8 mL/mol, respectively. As noted above, this behavior is driven by the diminishing population of isolated, fully hydrated ions. By contrast, the emergence of higher-order correlations is reflected in the growth of 3- and 4-body cluster contributions beyond 6.7 M, which increase by 10.6 and 8.0 mL/mol, respectively. These trends indicate that, at elevated concentrations, the electrolyte undergoes a pronounced structural reorganization involving clusters of all sizes.

The Voronoi tesselation approach described above underscores the fact that a molecular-scale decomposition is not accessible from the magnitudes, signs, or even cumulative contributions of the virial coefficients (B$_1$–B$_4$) alone. This represents a serious limitation to interpreting the virial expansion without explicit structural resolution.

\subsection{D. Probing Hydrogen Bond Network through Raman Spectra}

The preceding analyzes reveal collective synergistic responses of cations and anions across the full concentration range. The substantial water contribution up to the $V_m^{LiCl}$ maximum at 6.7 M reflects electrostriction, but this contribution by itself does not directly quantify how the underlying hydrogen-bonding network evolves. Radial distribution functions from our simulations (Figure S3.8 in the SI) point to a distortion of the hydrogen-bonding structure with reduced orientational ordering between pairs of neighboring waters, partially compensated by the donation of hydrogen bonds to Cl$^-$ ions. These structural signatures suggest a reorganization of the hydrogen bonding, consistent with past studies\cite{intro_zhang_review, num_den_nguyen, raman_mcr_geissler}, that cannot easily be resolved from volumetric observables alone, motivating a direct spectroscopic probe of the hydrogen-bonding network.

Raman spectroscopy measurements were performed to explore the vibrational spectra of water in the OH stretch region (3000–3800 cm$^{-1}$). Specifically, changes in the frequency and intensity of OH stretch resonances reflect the restructuring of the hydrogen bonding network as the salt concentration is increased. As a general rule, interactions that strengthen hydrogen bonds cause the peak to red shift in frequency, whereas weak or broken hydrogen bonds result in blue shifts, closer to the gas-phase (vacuum) frequency\cite{raman_water_1_expt, raman_water_2_expt, raman_red_shift_buck, raman_red_shift_markus, raman_red_shift_skinner}.

\begin{figure}[ht!]
    \centering
    \includegraphics[width=\linewidth]{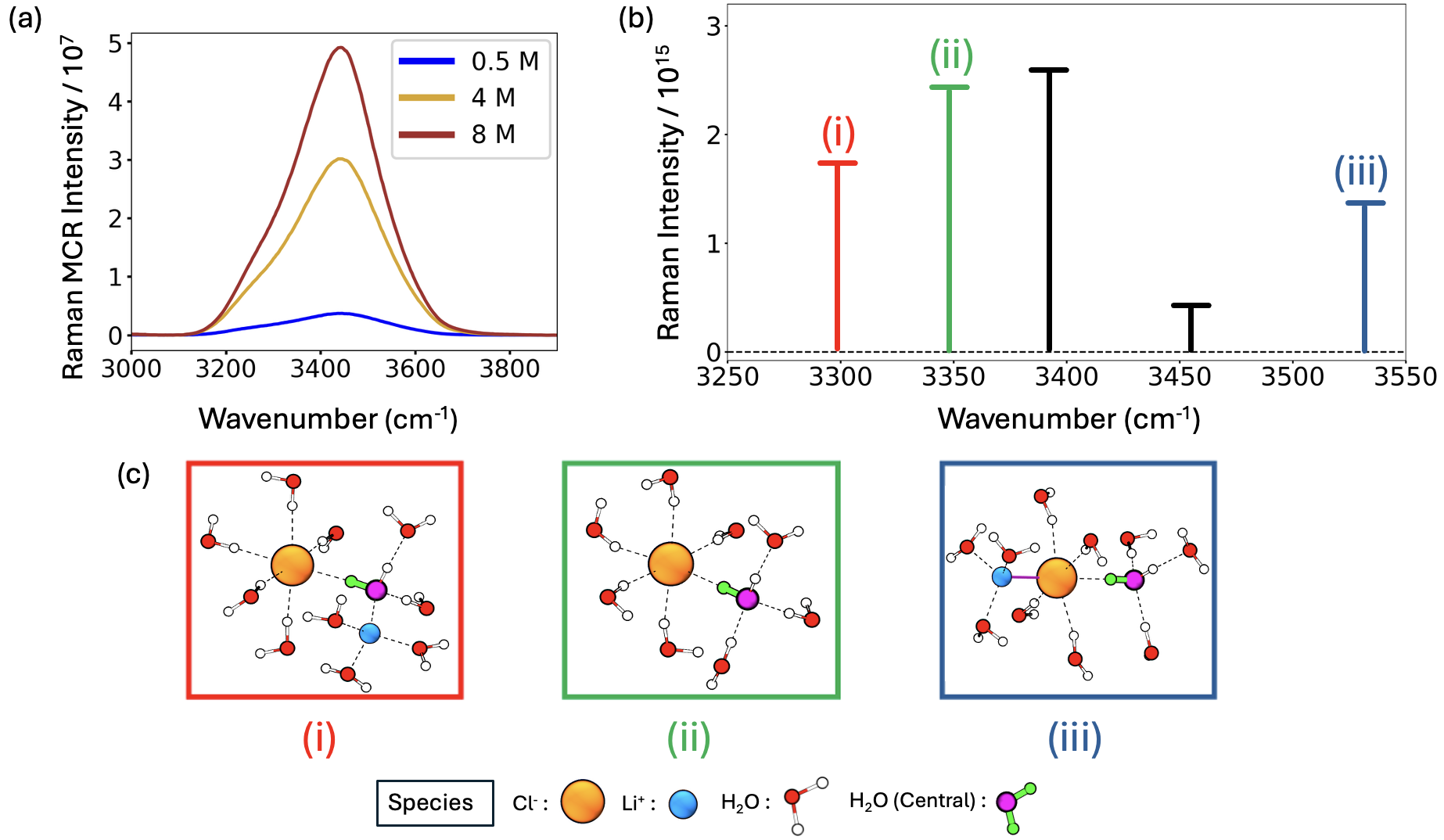}
    \caption{\textit{(a) Solute-correlated Raman data obtained by MCR at 0.5, 4.0 and 8.0 M. Note, additional isotopic dilution experiments were performed using a 1:9 H$_2$O/D$_2$O mixture. After decoupling the OH oscillators, the OH stretch region still gave rise to three distinct peaks upon performing MCR. These experiments confirmed that both the low and high frequency shoulders do not merely arise from intramolecular coupling of the OH stretch modes (see Section S2 in SI). (b) Raman spectrum from harmonic normal mode calculations of the clusters shown in panel (c). The black lines denote the OH stretch frequencies of pure water and are provided for reference. Peaks due to water molecules associated with other waters are divided into three categories: those with frequencies below 3330 cm$^{-1}$ (red lines) those with frequencies in the 3330-3470 cm$^{-1}$ range (green lines), and those with frequencies above 3470 cm$^{-1}$ (blue lines). (c) Illustration of the clusters used to obtain the theoretical Raman spectra : (i) a solvent shared ion pair, (ii) a solvated chloride ion and (iii) a contact ion pair.}}
    \label{fig:raman}
\end{figure}

To isolate and quantify changes, Raman spectra were deconvoluted using the multivariate curve resolution (MCR) method\cite{raman_mcr_1_dor, raman_mcr_2_dor}. For clarity, the extracted solute-correlated spectral signatures for 0.5 M, 4.0 M, and 8.0 M solutions are presented in Figure \ref{fig:raman}a, while additional spectra over the entire concentration range are provided in the SI (Section S2.1.B). The OH stretch profiles obtained from MCR exhibit a nearly monotonic increase in the solute-correlated component across the full range. This trend directly reflects the growing population of perturbed water molecules in the hydration shell of the ions compared with unperturbed bulk water. In addition to the enhancement of the dominant central peak at $\sim$3440 cm$^{-1}$, spectral deconvolution of the solute-correlated signal into Gaussian components reveals two distinct shoulder features: one red-shifted to $\sim$3270 cm$^{-1}$ and the other blue-shifted to $\sim$3590 cm$^{-1}$.

In order to elucidate the physical origins of the spectral features from the MCR Raman data, we performed quantum-chemistry calculations on model ion-water clusters inspired by our preceding analysis. The abstracted Raman Resonances are provided in Figure 5b, and the corresponding structures are shown in Figure \ref{fig:raman}c (i)-(iii). Details on how these clusters were constructed can be found in Section S3.7B of the SI. In the absence of explicit dynamical effects and the full ion environment, these calculations were intended to capture trends rather than absolute values. We focused on three distinct structural motifs inspired by our ion clusters (color coded in red, green, and blue), which helped rationalize the concentration dependence of the spectra and the putative emergence of the three experimental components observed from their deconvolution. The water molecules that give rise to these spectral features are enlarged and their oxygens are colored violet in Figure \ref{fig:raman}. The specific OH bonds involved in the corresponding vibrational modes are shown in green.


The black lines shown in Figure \ref{fig:raman}b at 3383 and 3453 cm$^{-1}$ correspond to the symmetric and asymmetric OH stretch modes of bulk water, respectively. When salt was introduced, a series of resonances with frequencies ranging from 3330 to 3470 cm$^{-1}$ arose which corresponded to water molecules hydrating Cl$^-$ ions (represented by the vertical green line in Figure \ref{fig:raman}b). We therefore assigned the growth of the experimental peak at 3440 cm$^{-1}$ to the first hydration shell around the anions (see the illustration in Figure \ref{fig:raman}c (ii)) with the OH bond pointing toward the Cl$^-$ anion. This produces a calculated vibrational mode at 3347 cm$^{-1}$, consistent with previous studies\cite{raman_mcr_geissler}. 

As the electrolyte concentration increased, ion pairing became a dominant structural feature, as established above through the PMV study. Some of the simplest consequential motifs: water molecules sandwiched between Li$^+$ and Cl$^-$ ion pairs (solvent shared ion pairs - see Figure \ref{fig:raman}c (i)) generally resulted in the most red-shifted calculated OH resonances with frequencies ranging between 3100-3300 cm$^{-1}$ (red vertical line in Figure \ref{fig:raman}b). This pronounced red shift can be understood in terms of a cooperative interaction between the cation and anion with Li$^+$ acting as a Lewis acid, enhancing the hydrogen-bond–donating ability of the sandwiched water molecule, while Cl$^-$ gave rise to a strong hydrogen-bond–accepting interaction. The combined effect substantially weakened the covalent OH stretch bond, leading to the experimentally observed red shift at 3270 cm$^{-1}$.


Finally, a Raman active mode close to the experimental peak at 3590 cm$^{-1}$ led to the identification of a cluster where water molecules solvated a contact ion pair (see Figure \ref{fig:raman}c (iii)). More specifically, it was found that water molecules solvating Cl$^-$ involved in contact ion pairing with Li$^+$ contributed to a calculated vibrational signature near 3532 cm$^{-1}$ (Figure 5b). The presence of Li$^+$ in direct contact with Cl$^-$ effectively reduced its ability to accept a hydrogen-bond from surrounding water molecules.


\section{Discussion}

\subsection{A. Electrolyte Models}

Several theoretical frameworks have been formulated to interpret the virial expansions typically observed in thermodynamic measurements of electrolyte properties. For example, McMillan-Mayer (MM) theory is a statistical mechanical model that can be applied to liquids with dissolved solutes\cite{mac_mayer_1, mac_mayer_2}. MM integrates out the solvent and treats the remaining solutes as if they were gas molecules confined inside a container. In this model, ion interactions are described by a potential of mean force (PMF) that averages over all possible solvent configurations. This approach is only valid at low salt concentrations. As such, it is difficult to employ MM to glean molecular-level insights into ion structures and clustering. Instead, investigators usually turn to Kirkwood-Buff integration\cite{kb_1, kb_2,smith2025,nico2025} (KBI), which is capable of describing two-body effects using RDFs. Unfortunately, converging KBI faces challenges for non-ideal systems where ion-clustering can lead to inhomogeneities in the system\cite{vegt2018}.

In addition to rigorous theories, empirical models have been developed to predict the behavior of electrolyte solutions. This includes work by Pitzer\cite{pitzer_eqn_1, pitzer_eqn_2} who employed a Debye-H\"uckel term to describe electrostatic interactions and screening at low salt concentrations in addition to a virial expansion. Curiously, the Pitzer model sets the linear term (B$_1$ in our notation) equal to zero and therefore describes pairwise interactions between ions strictly with the B$_2$ coefficient. Unfortunately, eliminating the linear term discards the key role of competitive water electrostriction in changing $V_m^{salt}$. Another more recent empirical model is the three-characteristic parameter correlation\cite{tcpc_1, tcpc_2} (TCPC). However, the performance of TCPC has typically been found to be poorer than Pitzer theory\cite{tcpc_pitzer_comp}. Also, Schwaab and Pezzotti have recently developed a theory of electrolyte solutions where the volume-environment interactions are incorporated through a generalized multipole expansion\cite{schwaabpezzotti2025}.

By contrast with previous theories and methods, the present framework treats the PMV as a molecular observable that directly reflects the collective structural organization of ions in electrolyte solutions. As described above, this method works by experimentally obtaining $V_m^{salt}$ over a wide range of salt concentrations. These values are then used to fine tune force fields for all-atom MD simulations. Even by employing this procedure, the power series contributions to $V_m^{salt}$ (Figure \ref{fig:vmol}a) are still challenging to interpret because ion-ion interactions involve a mixture of local and nonlocal contributions. For example, it is difficult to elucidate the relative contributions of solvent separated, solvent shared, and contact ion interactions in the B$_n$ terms. This obstacle to abstracting chemically useful information can be circumvented by using conformations generated with MD simulations and defining isolated ion, two-body, three-body, and four-body interactions strictly in terms of direct contact interactions. Figure \ref{fig:vmol}c does precisely that. From this construction, it is straightforward to visualize the evolution of ion clustering as a function of salt concentration. 

Defining ion clusters through contact interactions is not a unique choice. In fact, there should be numerous useful definitions that can be employed to understand electrolyte behavior. For example, one could place solvent shared, solvent separated and contact pairing interactions into separate categories\cite{allenmorita2023,craig2024}. This might even be a judicious means of unraveling the relative contributions from each category that lead to the B$_n$ terms shown in Figure \ref{fig:vmol}a. In this regard, the use of unsupervised learning approaches leveraging local-atomic descriptors\cite{ML_nikhil,ML_kcl_zhang,disc_giulia, disc_aq, disc_zundeig} to identify complex patterns of ions and water molecules would be an interesting avenue to explore in the future.


\subsection{B. The Evolution of Ion Clustering}

Figure \ref{fig:vmol}c is a useful starting point for elucidating the molecular-level details of many-body interactions. For example, clusters of size 3 (Agg 3) consist of two different species: LiCl$_2^-$ and Li$_2$Cl$^+$ (Figure 6a, top). These aggregates not only have opposite charges, but our MD simulations reveal that the anionic cluster is more abundant than the cationic one (see Figure S3.9 in the SI). The origin of this asymmetry is likely rooted in the fact that Cl$^-$ accepts more oriented hydrogen bonds compared to Li$^+$ leading to more charge-dipole interactions for the anionic cluster. By contrast, clusters of size four are predominantly neutral (Li$_2$Cl$_2$, Figure 6a, bottom), while the other four possible combinations (Li$_3$Cl$^{2+}$, LiCl$_3^{2-}$, Li$_4^{4+}$ or Cl$_4^{4-}$) are found with very small probability (Figure S3.9 in the SI). Both Agg 3 and Agg 4 clusters become increasingly important at higher concentrations (Figure 6b).  

As can be seen, the populations of the cationic and anionic Agg 3 cluster are both individually larger than the combined populations of all Agg 4 clusters (chains + rings). This is significant as the thermodynamic data in Figure 4a clearly shows that B$_4$[LiCl]$^4$ becomes more significant than B$_3$[LiCl]$^3$ at higher salt concentrations. As such, looking at the population of Agg 4 made strictly from contact ion interactions underweights 4-body interactions while simultaneously overweighting 3-body interactions.  

\begin{figure}
    \centering
    \includegraphics[width=\linewidth]{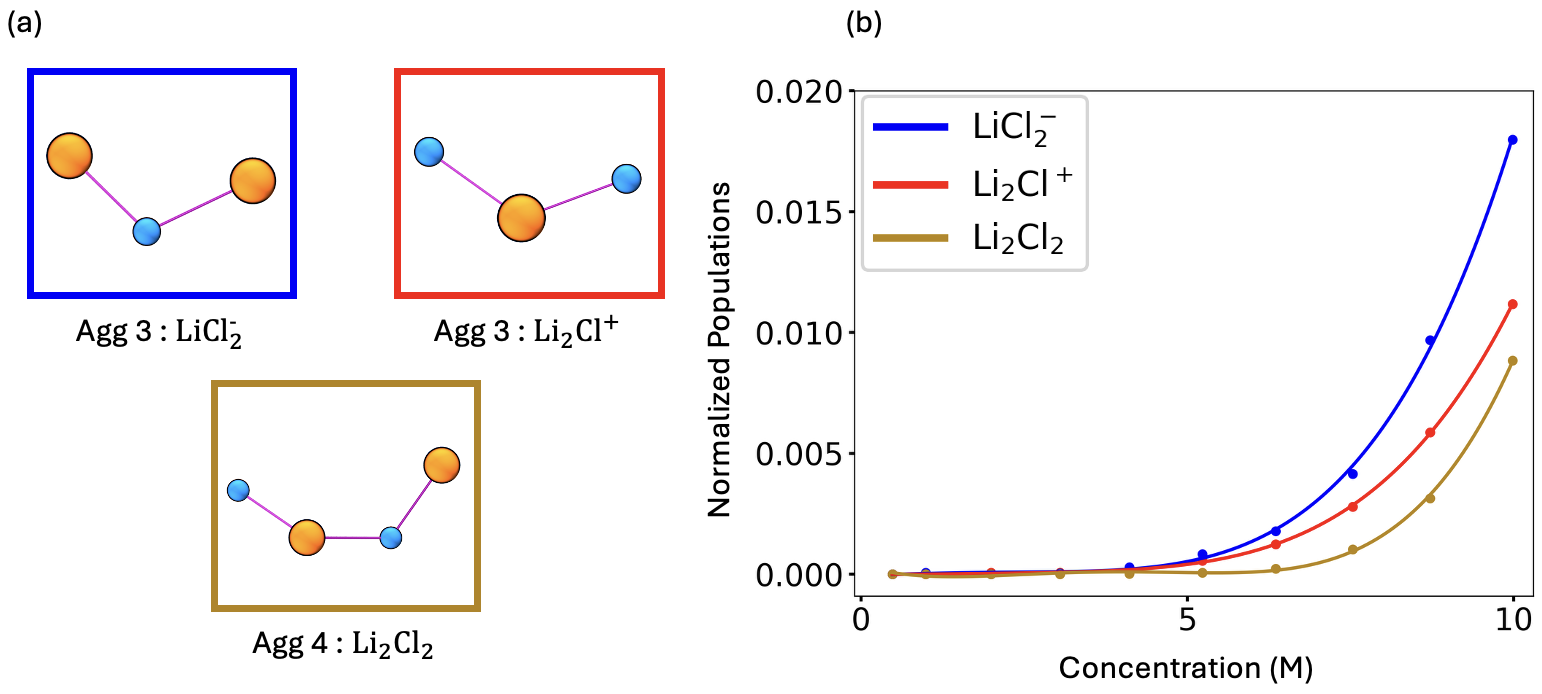}
    \caption{\textit{(a) Example of clusters of formula unit : LiCl$_2^-$, Li$_2$Cl$^+$ and Li$_2$Cl$_2$ and (b) The evolution of the time averaged populations of the corresponding species with concentration. Populations are normalized by the total number of ions at the respective concentrations.}}
    \label{fig:pop}
\end{figure}

As noted in the introduction, accurate $V_m^{salt}$ and $V_m^{H_2O}$ curves were not available before the current study\cite{partial_lamer}. The reason for this is two-fold. First, most density measurements\cite{apparant_density_wolf, apparant_density_sohnel_novotny, apparant_density_klugman} of electrolyte solutions made in the 20$^{\text{th}}$ century did not have sufficient precision and accuracy to produce reliable PMV curves.  Second, one needs to employ density data in combination with Equations \ref{eq:form_vm_salt} and \ref{eq:form_vm_wat} rather than using procedures that yield the apparent PMV\cite{apparant_baxter, apparant_green, apparant_masson, apparant_nernst, apparant_owen, apparant_pitzer, apparant_vaslow, apparent_vercher} (see Section S1 in the SI for further discussion). 

With proper curves in hand, it is possible to compare the data in Figures \ref{fig:dens}b and  \ref{fig:dens}c to a wide variety of thermodynamic properties for LiCl solutions. Remarkably, the shape of the $V_m^{H_2O}$ curve is reminiscent of data for the freezing point depression of LiCl solutions as a function of salt concentration (see Figure S1.6 in the SI). In particular, the eutectic point\cite{eutectic_conde} occurs at 6.8 M, which is quite close to the minimum in $V_m^{H_2O}$ near 6.7 M, despite the fact that the eutectic point is roughly 100$^\circ$C below the temperature at which density measurements are made in the current study (20$^\circ$C). This fact suggests that the organization of ion clusters is only weakly dependent on temperature. Moreover, the eutectic point and PMV minimum/maximum occur almost exactly at the same salt concentration where the chemical potential of water falls most rapidly as determined by vapor pressure osmotic (VPO) studies\cite{vpo_pearce} (see Figure S1.6 in the SI). 

The interconnected nature of PMV, VPO, and eutectic point measurements suggests that 6.7 M LiCl represents a pivotal salt concentration. In fact, this is also near the concentration where the second derivative of the cluster size 2 (Agg 2) curve reaches a maximum  (see Figure S3.12 in the SI). In addition, clusters of size three and four (Agg 3 and Agg 4) become significantly more prominent above this point. Moreover, the volume contraction of water molecules shrinks until 6.7 M but has a constant contribution to $V_m^{LiCl}$ at higher salt concentrations (Figure \ref{fig:vmol}c). Taken together, it appears that the macroscopic thermodynamic properties of the solution are closely linked to microscopic ion pairing behavior at this concentration.

\begin{figure}
    \centering    
    \includegraphics[width=\linewidth]{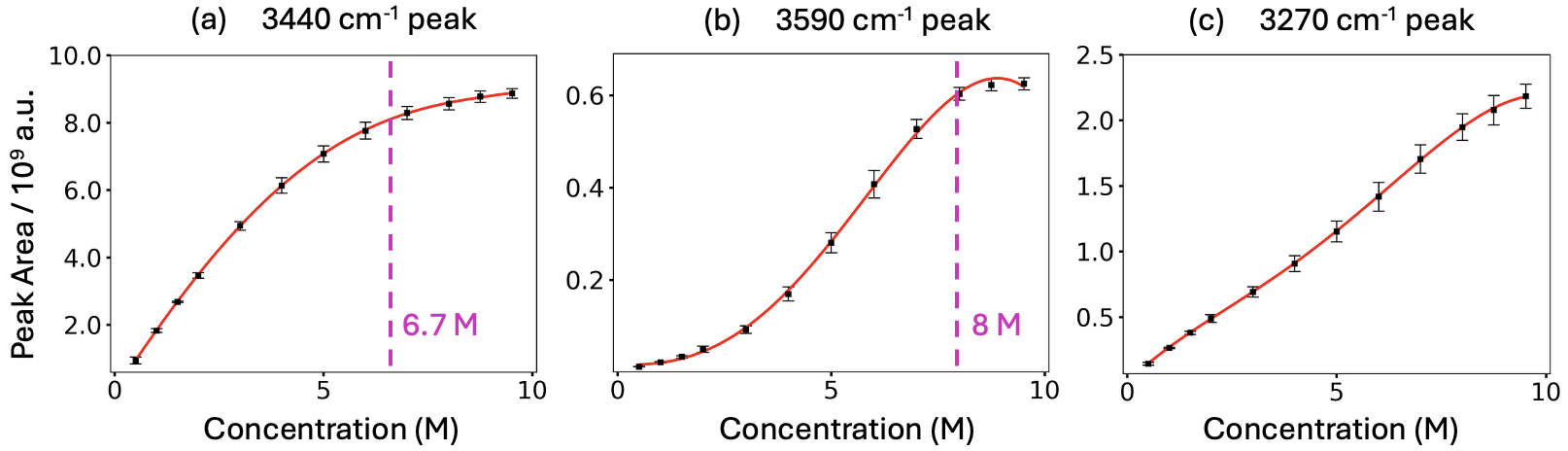}
    \caption{\textit{Evolution of the area under the Gaussian components resolved from the solute-correlated part of the Raman signal centered at (a) 3440 cm$^{-1}$, (b) 3590 cm$^{-1}$, and (c) 3270 cm$^{-1}$.}}
    \label{fig:peak}
\end{figure}

Next, the Raman MCR spectra in the OH stretch region confirm the idea water behaves differently beyond 6.7 M LiCl. Specifically, the area under the 3440 cm$^{-1}$ peak continuously increases until 6.7 M, but levels off rapidly past this point (Figure \ref{fig:peak}a). This is direct spectroscopic evidence that complete first hydration shells form around isolated Cl$^-$ species at lower concentrations (Figure \ref{fig:raman}c (ii)). However, sufficient water is no longer available beyond this point to accommodate complete hydration of additional salt ions. 

The 3590 cm$^{-1}$ resonance follows a similar trend to the 3440 cm$^{-1}$ peak but does not begin to level off until a somewhat higher salt concentration ($>$ 8 M LiCl, Figure \ref{fig:peak}b). This higher frequency peak represents OH groups that point toward Cl$^-$ ions that are in contact pair configurations with Li$^+$ on the opposite side of the anion (Figure 5c (iii)). As such, the electron density around Cl$^-$ is distorted by Li$^+$, which attenuates the charge density that can be transferred from the anion into the $\sigma^*$ orbital of water molecules that are hydrogen bonded to it\cite{hbond_ct_herbert}. Moreover, the increase in the 3590 cm$^{-1}$ peak area roughly follows the shape of the Agg 2 cluster curve (Figure \ref{fig:vmol}c), which also levels off near 8 M LiCl. By contrast, the 3270 cm$^{-1}$ resonance continuously grows from low LiCl concentration to 9.5 M (Figure \ref{fig:peak}c). This is consistent with the fact that the concentration of solvent shared ion pairs continuously increases throughout the entire concentration range (Figure \ref{fig:raman}c (i)). At relatively low salt concentrations, water molecules that interact with both a cation and an anion, largely represent solvent-shared ion pairs without other ions in the vicinity. At higher salt concentrations, these same water molecules are increasingly shared between adjacent chains and rings.

\hspace{0.3 cm}
\hspace{0.3 cm}

\subsection{C. Toward Reliable Electrolyte Models}

The refinement of classical empirical potentials for modeling electrolyte solutions is an active area of current research. Specifically, the scaling of charges on ions is considered to be an important means for modeling electronic screening effects. We show that LJ parameters can be optimized for both scaled and unscaled charges to reproduce experimental PMV values. This provides an option to compensate for very different partial charges consistent with recent reports by Jungwirth and co-workers\cite{ions_pavel_2025}.  However, simultaneously reproducing surface tension, viscosity, VPO, conductivity, and PMV measurements across diverse salts and concentrations may require the use of a fixed charge scaling value. Moreover, as more precise electrolyte work is done, relaxation of the Lennard–Jones combination rules and/or the incorporation of additional functional forms, such as empirical charge–dipole interactions, may be required. However, it is becoming clearer that highly accurate experimental measurements are a necessary prerequisite for generating reliable ion force fields for use in simulations. Specifically, the use of 20$^{\text{th}}$ century data for parameterizing ion force fields in water should only be done with great caution.

Beyond the optimization of force fields, renewed attention to the generation of improved electrolyte models should have broad ramifications for understanding ion-specific phenomena across chemistry, biology, and materials science. Indeed, the Hofmeister series has been known for 138 years, yet the underlying molecular-level reasons for the recurring ranking of ions across a wide variety of physical phenomena still remains the subject of intense debate. We suggest that the lack of high quality electrolyte data represents a crucial reason why Hofmeister chemistry has not yet been better understood. In fact, deeper insight will require the continued integration of thermodynamics, spectroscopy, and molecular simulation for electrolyte studies, as exemplified by the present work.




\section{Acknowledgments}

C.T.L. and P.S.C. thank the National Science Foundation (CHE-2305129 and CHE-2154651) for support. D.D. and A.H. acknowledge funding from the European Research Council (ERC) under the European Union’s Horizon 2020 research and innovation program (grant agreement No. 101043272 – HyBOP). The views and opinions expressed are those of the authors only and do not necessarily reflect those of the European Union or the European Research Council Executive Agency. Neither the European Union nor the granting authority can be held responsible for them. D.D and A.H. also acknowledge MareNostrum5 (project EHPC-EXT-2023E01-029) for computational resources.


\bibliography{LiCl}

@article{hbond_ct_herbert,
  title={Electrostatics, charge transfer, and the nature of the halide--water hydrogen bond},
  author={Herbert, John M and Carter-Fenk, Kevin},
  journal={The Journal of Physical Chemistry A},
  volume={125},
  number={5},
  pages={1243--1256},
  year={2021},
  publisher={ACS Publications}
}

@article{num_den_nguyen,
	author = {Nguyen, Man Thi Hong and Tichacek, Ondrej and Martinez-Seara, Hector and Mason, Philip E. and Jungwirth, Pavel},
	date = {2021/04/01},
	doi = {10.1021/acs.jpcb.0c10599},
	isbn = {1520-6106},
	journal = {The Journal of Physical Chemistry B},
	journal1 = {The Journal of Physical Chemistry B},
	journal2 = {J. Phys. Chem. B},
	month = {04},
	number = {12},
	pages = {3153--3162},
	publisher = {American Chemical Society},
	title = {Resolving the Equal Number Density Puzzle: Molecular Picture from Simulations of LiCl(aq) and NaCl(aq)},
	type = {doi: 10.1021/acs.jpcb.0c10599},
	url = {https://doi.org/10.1021/acs.jpcb.0c10599},
	volume = {125},
	year = {2021},
	year1 = {2021}}

@article{vegt2018,
	author = {Milzetti, Jasmin and Nayar, Divya and van der Vegt, Nico F. A.},
	doi = {10.1021/acs.jpcb.7b11831},
	eprint = {https://doi.org/10.1021/acs.jpcb.7b11831},
	journal = {The Journal of Physical Chemistry B},
	note = {PMID: 29342355},
	number = {21},
	pages = {5515-5526},
	title = {Convergence of Kirkwood--Buff Integrals of Ideal and Nonideal Aqueous Solutions Using Molecular Dynamics Simulations},
	url = {https://doi.org/10.1021/acs.jpcb.7b11831},
	volume = {122},
	year = {2018}}

@article{schwaabpezzotti2025,
	author = {Schwaab, Gerhard and Pezzotti, Simone},
	doi = {10.1039/D5CP01781E},
	issue = {33},
	journal = {Phys. Chem. Chem. Phys.},
	pages = {17585-17597},
	publisher = {The Royal Society of Chemistry},
	title = {A new perspective on aqueous electrolyte solutions},
	url = {http://dx.doi.org/10.1039/D5CP01781E},
	volume = {27},
	year = {2025}}

@article{ML_kcl_zhang,
  title={Dissolving salt is not equivalent to applying a pressure on water},
  author={Zhang, Chunyi and Yue, Shuwen and Panagiotopoulos, Athanassios Z and Klein, Michael L and Wu, Xifan},
  journal={Nature communications},
  volume={13},
  number={1},
  pages={822},
  year={2022},
  publisher={Nature Publishing Group UK London}
}

@article{ML_nikhil,
  title={Understanding the anomalous diffusion of water in aqueous electrolytes using machine learned potentials},
  author={Avula, Nikhil VS and Klein, Michael L and Balasubramanian, Sundaram},
  journal={The Journal of Physical Chemistry Letters},
  volume={14},
  number={42},
  pages={9500--9507},
  year={2023},
  publisher={ACS Publications}
}

@article{nico2025,
    author = {Chattopadhyay, Abhishek and Mandalaparthy, Varun and van der Vegt, Nico F. A.},
    title = {Determination of aqueous solubility of NaCl in molecular dynamics simulation using the Kirkwood–Buff method},
    journal = {The Journal of Chemical Physics},
    volume = {162},
    number = {17},
    pages = {174116},
    year = {2025},
    month = {05},
    issn = {0021-9606},
    doi = {10.1063/5.0264104},
    url = {https://doi.org/10.1063/5.0264104},
    eprint = {https://pubs.aip.org/aip/jcp/article-pdf/doi/10.1063/5.0264104/20510717/174116_1_5.0264104.pdf},
}

@article{allenmorita2023,
	author = {Wang, Lin and Morita, Akihiro and North, Nicole M. and Baumler, Stephen M. and Springfield, Elliot W. and Allen, Heather C.},
	doi = {10.1021/acs.jpcb.2c07923},
	eprint = {https://doi.org/10.1021/acs.jpcb.2c07923},
	journal = {The Journal of Physical Chemistry B},
	note = {PMID: 36757371},
	number = {7},
	pages = {1618-1627},
	title = {Identification of Ion Pairs in Aqueous NaCl and KCl Solutions in Combination with Raman Spectroscopy, Molecular Dynamics, and Quantum Chemical Calculations},
	url = {https://doi.org/10.1021/acs.jpcb.2c07923},
	volume = {127},
	year = {2023}}

@article{craig2024,
	author = {Elliott, Gareth R. and Wanless, Erica J. and Webber, Grant B. and Andersson, Gunther G. and Craig, Vincent S. J. and Page, Alister J.},
	doi = {10.1021/acs.jpcb.4c01992},
	eprint = {https://doi.org/10.1021/acs.jpcb.4c01992},
	journal = {The Journal of Physical Chemistry B},
	note = {PMID: 39037039},
	number = {30},
	pages = {7438-7444},
	title = {Dynamic Ion Correlations and Ion-Pair Lifetimes in Aqueous Alkali Metal Chloride Electrolytes},
	url = {https://doi.org/10.1021/acs.jpcb.4c01992},
	volume = {128},
	year = {2024}}

@article{smith2025,
	author = {Ploetz, Elizabeth A. and Smyers, Nathan D. and Smith, Paul E.},
	doi = {10.1021/acs.jpcb.4c07583},
	eprint = {https://doi.org/10.1021/acs.jpcb.4c07583},
	journal = {The Journal of Physical Chemistry B},
	note = {PMID: 39817653},
	number = {4},
	pages = {1387-1398},
	title = {Ion--Ion Association in Bulk Mixed Electrolytes Using Global and Local Electroneutrality Constraints},
	url = {https://doi.org/10.1021/acs.jpcb.4c07583},
	volume = {129},
	year = {2025}}

@article{fayer2022,
	author = {Roget, Sean A. and Carter-Fenk, Kimberly A. and Fayer, Michael D.},
	doi = {10.1021/jacs.2c00616},
	eprint = {https://doi.org/10.1021/jacs.2c00616},
	journal = {Journal of the American Chemical Society},
	note = {PMID: 35226487},
	number = {9},
	pages = {4233-4243},
	title = {Water Dynamics and Structure of Highly Concentrated LiCl Solutions Investigated Using Ultrafast Infrared Spectroscopy},
	url = {https://doi.org/10.1021/jacs.2c00616},
	volume = {144},
	year = {2022}}

@article{basis_3,
	author = {Kostal, Vojtech and Mason, Philip E and Martinez-Seara, Hector and Jungwirth, Pavel},
	journal = {The Journal of Physical Chemistry Letters},
	number = {19},
	pages = {4403--4408},
	publisher = {ACS Publications},
	title = {Common cations are not polarizable: Effects of dispersion correction on hydration structures from ab initio molecular dynamics},
	volume = {14},
	year = {2023}}

@article{ions_pavel_2025,
	author = {Fan, Shujie and Mason, Philip E and Chamorro, Victor Cruces and Shanks, Brennon L and Martinez-Seara, Hector and Jungwirth, Pavel},
	journal = {Journal of Chemical Theory and Computation},
	number = {18},
	pages = {9023--9034},
	publisher = {ACS Publications},
	title = {Charge scaling force field for biologically relevant ions utilizing a global optimization method},
	volume = {21},
	year = {2025}}

@article{disc_zundeig,
	author = {Di Pino, Solana and Donkor, Edward Danquah and S{\'a}nchez, Veronica M and Rodriguez, Alex and Cassone, Giuseppe and Scherlis, Damian and Hassanali, Ali},
	journal = {The Journal of Physical Chemistry B},
	number = {45},
	pages = {9822--9832},
	publisher = {ACS Publications},
	title = {Zundeig: The structure of the proton in liquid water from unsupervised learning},
	volume = {127},
	year = {2023}}

@article{disc_aq,
	author = {Banerjee, Debarshi and Azizi, Khatereh and Egan, Colin K and Donkor, Edward Danquah and Malosso, Cesare and Pino, Solana Di and Mir{\'o}n, Gonzalo D{\'\i}az and Stella, Martina and Sormani, Giulia and Hozana, Germaine Neza and others},
	journal = {Chemical Physics Reviews},
	number = {2},
	publisher = {AIP Publishing},
	title = {Aqueous solution chemistry in silico and the role of data-driven approaches},
	volume = {5},
	year = {2024}}

@book{rdf_allen_tildsley,
	author = {Allen, Michael P and Tildesley, Dominic J},
	publisher = {Oxford university press},
	title = {Computer simulation of liquids, chapter 2},
	year = {2017}}

@article{partial_lamer,
	author = {LaMer, V. K. and Gronwall, T. H.},
	date = {1927/03/01},
	doi = {10.1021/j150273a005},
	isbn = {0092-7325},
	journal = {The Journal of Physical Chemistry},
	journal1 = {The Journal of Physical Chemistry},
	journal2 = {J. Phys. Chem.},
	month = {03},
	number = {3},
	pages = {393--406},
	publisher = {American Chemical Society},
	title = {The Partial Molal Volumes of Water and Salt in Solutions of the Alkali Halides},
	type = {doi: 10.1021/j150273a005},
	url = {https://doi.org/10.1021/j150273a005},
	volume = {31},
	year = {1927},
	year1 = {1927}}

@article{gibbs_duhem_Gokcen,
	author = {Gokcen, Nev A. },
	date = {1960/04/01},
	doi = {10.1021/j100833a006},
	isbn = {0022-3654},
	journal = {The Journal of Physical Chemistry},
	journal1 = {The Journal of Physical Chemistry},
	journal2 = {J. Phys. Chem.},
	month = {04},
	number = {4},
	pages = {401--406},
	publisher = {American Chemical Society},
	title = {Application of Gibbs and GIBBS—DUHEM Equations to Ternary and Multicomponent Systems},
	type = {doi: 10.1021/j100833a006},
	url = {https://doi.org/10.1021/j100833a006},
	volume = {64},
	year = {1960},
	year1 = {1960}}

@article{intro_li_su,
	author = {Su, Yuan and Ryder, John and Li, Baolin and Wu, Xin and Fox, Niles and Solenberg, Pat and Brune, Kellie and Paul, Steven and Zhou, Yan and Liu, Feng and Ni, Binhui},
	date = {2004/06/01},
	doi = {10.1021/bi035627j},
	isbn = {0006-2960},
	journal = {Biochemistry},
	journal1 = {Biochemistry},
	journal2 = {Biochemistry},
	month = {06},
	number = {22},
	pages = {6899--6908},
	publisher = {American Chemical Society},
	title = {Lithium, a Common Drug for Bipolar Disorder Treatment, Regulates Amyloid-$\beta$Precursor Protein Processing},
	type = {doi: 10.1021/bi035627j},
	url = {https://doi.org/10.1021/bi035627j},
	volume = {43},
	year = {2004},
	year1 = {2004}}

@article{intro_li_alzheimer,
	author = {Aron, Liviu and Ngian, Zhen Kai and Qiu, Chenxi and Choi, Jaejoon and Liang, Marianna and Drake, Derek M. and Hamplova, Sara E. and Lacey, Ella K. and Roche, Perle and Yuan, Monlan and Hazaveh, Saba S. and Lee, Eunjung A. and Bennett, David A. and Yankner, Bruce A.},
	date = {2025/09/01},
	doi = {10.1038/s41586-025-09335-x},
	id = {Aron2025},
	isbn = {1476-4687},
	journal = {Nature},
	number = {8081},
	pages = {712--721},
	title = {Lithium deficiency and the onset of Alzheimer's disease},
	url = {https://doi.org/10.1038/s41586-025-09335-x},
	volume = {645},
	year = {2025}}

@book{intro_lehninger,
	author = {Nelson, D.L. and Lehninger, A.L. and Cox, M.M.},
	isbn = {9780716771081},
	lccn = {2007941224},
	publisher = {W. H. Freeman},
	series = {Lehninger Principles of Biochemistry},
	title = {Lehninger Principles of Biochemistry, chapter 11},
	url = {https://books.google.it/books?id=5Ek9J4p3NfkC},
	year = {2008}}

@article{intro_carboncapture_hegarty,
	author = {Hegarty, John and Shindel, Benjamin and Sukhareva, Daria and Barsoum, Michael L. and Farha, Omar K. and Dravid, Vinayak},
	date = {2023/12/19},
	doi = {10.1021/acs.est.3c02543},
	isbn = {0013-936X},
	journal = {Environmental Science \& Technology},
	journal1 = {Environmental Science \& Technology},
	journal2 = {Environ. Sci. Technol.},
	month = {12},
	number = {50},
	pages = {21080--21091},
	publisher = {American Chemical Society},
	title = {Expanding the Library of Ions for Moisture-Swing Carbon Capture},
	type = {doi: 10.1021/acs.est.3c02543},
	url = {https://doi.org/10.1021/acs.est.3c02543},
	volume = {57},
	year = {2023},
	year1 = {2023}}

@article{intro_wastewater_nemes,
	address = {Research Institute for Renewable Energies-ICER, Politehnica University Timisoara, Gavril Musicescu Street No. 138, 300774 Timisoara, Romania.; Faculty of Chemical Engineering, Biotechnologies and Environmental Protection, Politehnica University Timisoara, Victories Square No. 2, 300006 Timisoara, Romania.; Faculty of Chemical Engineering, Biotechnologies and Environmental Protection, Politehnica University Timisoara, Victories Square No. 2, 300006 Timisoara, Romania.; Faculty of Chemical Engineering, Biotechnologies and Environmental Protection, Politehnica University Timisoara, Victories Square No. 2, 300006 Timisoara, Romania.; Faculty of Chemical Engineering, Biotechnologies and Environmental Protection, Politehnica University Timisoara, Victories Square No. 2, 300006 Timisoara, Romania.; Department of Cardiology, Victor Babes University of Medicine and Pharmacy Timisoara, 2 Eftimie Murgu Square No. 2, 300041 Timisoara, Romania.},
	auid = {ORCID: 0000-0003-2532-2893; ORCID: 0000-0001-7572-3703; ORCID: 0000-0002-5015-9306},
	author = {Nemeș, Nicoleta Sorina and Negrea, Adina and Ciopec, Mihaela and Negrea, Petru and Du{\c t}eanu, Narcis and Duda-Seiman, Daniel Marius},
	cois = {The authors declare no conflicts of interest.},
	crdt = {2026/02/27 01:09},
	date = {2026 Feb 11},
	dcom = {20260306},
	dep = {20260211},
	doi = {10.3390/ijms27041741},
	edat = {2026/02/27 06:31},
	gr = {RORS00063/INTERREG IPA Romania-Serbia Program/},
	issn = {1422-0067 (Electronic); 1422-0067 (Linking)},
	jid = {101092791},
	journal = {Int J Mol Sci},
	jt = {International journal of molecular sciences},
	language = {eng},
	lid = {10.3390/ijms27041741 {$[$}doi{$]$}; 1741},
	lr = {20260306},
	mh = {*Metals, Heavy/isolation \& purification/chemistry; *Wastewater/chemistry; *Water Purification/methods; *Water Pollutants, Chemical/isolation \& purification/chemistry; Humans; Ions},
	mhda = {2026/03/06 19:43},
	month = {Feb},
	number = {4},
	oto = {NOTNLM},
	own = {NLM},
	phst = {2026/01/09 00:00 {$[$}received{$]$}; 2026/02/07 00:00 {$[$}revised{$]$}; 2026/02/09 00:00 {$[$}accepted{$]$}; 2026/03/06 19:43 {$[$}medline{$]$}; 2026/02/27 06:31 {$[$}pubmed{$]$}; 2026/02/27 01:09 {$[$}entrez{$]$}; 2026/02/11 00:00 {$[$}pmc-release{$]$}},
	pii = {ijms27041741; ijms-27-01741},
	pl = {Switzerland},
	pmc = {PMC12940842},
	pmcr = {2026/02/11},
	pmid = {41751874},
	pst = {epublish},
	pt = {Journal Article; Review},
	rn = {0 (Metals, Heavy); 0 (Wastewater); 0 (Water Pollutants, Chemical); 0 (Ions)},
	sb = {IM},
	status = {MEDLINE},
	title = {Heavy Metal Ion Removal: A Global Review of Wastewater Treatment Technologies.},
	volume = {27},
	year = {2026}}

@article{intro_protein_sol_kim,
	author = {Kim D. Collins},
	doi = {https://doi.org/10.1016/j.ymeth.2004.03.021},
	issn = {1046-2023},
	journal = {Methods},
	note = {Macromolecular Crystallization},
	number = {3},
	pages = {300-311},
	title = {Ions from the Hofmeister series and osmolytes: effects on proteins in solution and in the crystallization process},
	url = {https://www.sciencedirect.com/science/article/pii/S1046202304001124},
	volume = {34},
	year = {2004}}

@article{disc_giulia,
	author = {Sormani, Giulia and Rodriguez, Alex and Hassanali, Ali},
	date = {2025/08/26},
	doi = {10.1021/acs.jctc.5c00449},
	isbn = {1549-9618},
	journal = {Journal of Chemical Theory and Computation},
	journal1 = {Journal of Chemical Theory and Computation},
	journal2 = {J. Chem. Theory Comput.},
	month = {08},
	number = {16},
	pages = {8060--8072},
	publisher = {American Chemical Society},
	title = {Opportunities and Challenges in Unsupervised Learning: The Case of Aqueous Electrolyte Solutions},
	type = {doi: 10.1021/acs.jctc.5c00449},
	url = {https://doi.org/10.1021/acs.jctc.5c00449},
	volume = {21},
	year = {2025},
	year1 = {2025}}

@article{intro_solvent_extract,
	author = {Dupont, David and Depuydt, Daphne and Binnemans, Koen},
	date = {2015/06/04},
	doi = {10.1021/acs.jpcb.5b02980},
	isbn = {1520-6106},
	journal = {The Journal of Physical Chemistry B},
	journal1 = {The Journal of Physical Chemistry B},
	journal2 = {J. Phys. Chem. B},
	month = {06},
	number = {22},
	pages = {6747--6757},
	publisher = {American Chemical Society},
	title = {Overview of the Effect of Salts on Biphasic Ionic Liquid/Water Solvent Extraction Systems: Anion Exchange, Mutual Solubility, and Thermomorphic Properties},
	type = {doi: 10.1021/acs.jpcb.5b02980},
	url = {https://doi.org/10.1021/acs.jpcb.5b02980},
	volume = {119},
	year = {2015},
	year1 = {2015}}

@article{intro_corrosion_jiang,
	author = {Jiang, X and Zheng, YG and Qu, DR and Ke, W},
	journal = {Corrosion science},
	number = {10},
	pages = {3091--3108},
	publisher = {Elsevier},
	title = {Effect of calcium ions on pitting corrosion and inhibition performance in CO2 corrosion of N80 steel},
	volume = {48},
	year = {2006}}

@article{intro_corrosion_ma,
	author = {Ma, HY and Yang, C and Li, GY and Guo, WJ and Chen, SH and Luo, JL},
	journal = {Corrosion},
	number = {12},
	pages = {1112--1119},
	publisher = {Association for Materials Protection and Performance},
	title = {Influence of nitrate and chloride ions on the corrosion of iron},
	volume = {59},
	year = {2003}}

@article{intro_battery_bian,
	author = {Bian, Jiang and He, Hongyan and Liu, Yawei and Lai, Qiao and Zhao, Qian and Xu, Hui and Huo, Feng},
	journal = {Chemical Engineering Science},
	pages = {123257},
	publisher = {Elsevier},
	title = {Molecular insights into the effect of anion synergy on lithium-ion transport properties in highly concentrated dual-anion ionic liquid electrolytes},
	year = {2025}}

@article{intro_battery_suo,
	author = {Suo, Liumin and Borodin, Oleg and Gao, Tao and Olguin, Marco and Ho, Janet and Fan, Xiulin and Luo, Chao and Wang, Chunsheng and Xu, Kang},
	journal = {Science},
	number = {6263},
	pages = {938--943},
	publisher = {American Association for the Advancement of Science},
	title = {``Water-in-salt'' electrolyte enables high-voltage aqueous lithium-ion chemistries},
	volume = {350},
	year = {2015}}

@article{intro_kunz,
	author = {Kunz, Werner},
	journal = {Current Opinion in Colloid \& Interface Science},
	number = {1-2},
	pages = {34--39},
	publisher = {Elsevier},
	title = {Specific ion effects in colloidal and biological systems},
	volume = {15},
	year = {2010}}

@article{intro_zhang_review,
	author = {Zhang, Yanjie and Cremer, Paul S},
	journal = {Current opinion in chemical biology},
	number = {6},
	pages = {658--663},
	publisher = {Elsevier},
	title = {Interactions between macromolecules and ions: the Hofmeister series},
	volume = {10},
	year = {2006}}

@article{quantum_r2scan,
	author = {Furness, James W and Kaplan, Aaron D and Ning, Jinliang and Perdew, John P and Sun, Jianwei},
	journal = {The journal of physical chemistry letters},
	number = {19},
	pages = {8208--8215},
	publisher = {ACS Publications},
	title = {Accurate and numerically efficient r2SCAN meta-generalized gradient approximation},
	volume = {11},
	year = {2020}}

@article{basis_2,
	author = {Rappoport, Dmitrij and Furche, Filipp},
	journal = {The Journal of chemical physics},
	number = {13},
	publisher = {AIP Publishing},
	title = {Property-optimized Gaussian basis sets for molecular response calculations},
	volume = {133},
	year = {2010}}

@article{basis_1,
	author = {Weigend, Florian and Ahlrichs, Reinhart},
	journal = {Physical Chemistry Chemical Physics},
	number = {18},
	pages = {3297--3305},
	publisher = {Royal Society of Chemistry},
	title = {Balanced basis sets of split valence, triple zeta valence and quadruple zeta valence quality for H to Rn: Design and assessment of accuracy},
	volume = {7},
	year = {2005}}

@article{optim_li_1,
	author = {Li, Da-Wei and Bruschweiler, Rafael},
	journal = {Journal of chemical theory and computation},
	number = {6},
	pages = {1773--1782},
	publisher = {ACS Publications},
	title = {Iterative optimization of molecular mechanics force fields from NMR data of full-length proteins},
	volume = {7},
	year = {2011}}

@article{optim_hummer_3,
	author = {K{\"o}finger, J{\"u}rgen and Hummer, Gerhard},
	date = {2021/12/17},
	doi = {10.1140/epjb/s10051-021-00234-4},
	id = {K{\"o}finger2021},
	isbn = {1434-6036},
	journal = {The European Physical Journal B},
	number = {12},
	pages = {245},
	title = {Empirical optimization of molecular simulation force fields by Bayesian inference},
	url = {https://doi.org/10.1140/epjb/s10051-021-00234-4},
	volume = {94},
	year = {2021}}

@article{optim_hummer_1,
	author = {Hummer, Gerhard and K{\"o}finger, J{\"u}rgen},
	doi = {10.1063/1.4937786},
	eprint = {https://pubs.aip.org/aip/jcp/article-pdf/doi/10.1063/1.4937786/15510660/243150_1_online.pdf},
	issn = {0021-9606},
	journal = {The Journal of Chemical Physics},
	month = {12},
	number = {24},
	pages = {243150},
	title = {Bayesian ensemble refinement by replica simulations and reweighting},
	url = {https://doi.org/10.1063/1.4937786},
	volume = {143},
	year = {2015}}

@article{optim_hummer_2,
	author = {K{\"o}finger, J{\"u}rgen and Stelzl, Lukas S. and Reuter, Klaus and Allande, C{\'e}sar and Reichel, Katrin and Hummer, Gerhard},
	doi = {10.1021/acs.jctc.8b01231},
	eprint = {https://doi.org/10.1021/acs.jctc.8b01231},
	journal = {Journal of Chemical Theory and Computation},
	note = {PMID: 30939006},
	number = {5},
	pages = {3390-3401},
	title = {Efficient Ensemble Refinement by Reweighting},
	url = {https://doi.org/10.1021/acs.jctc.8b01231},
	volume = {15},
	year = {2019}}

@article{optim_bussi_3,
	author = {Gilardoni, Ivan and Fr{\"o}hlking, Thorben and Bussi, Giovanni},
	journal = {The Journal of Physical Chemistry Letters},
	number = {5},
	pages = {1204--1210},
	publisher = {ACS Publications},
	title = {Boosting ensemble refinement with transferable force-field corrections: Synergistic optimization for molecular simulations},
	volume = {15},
	year = {2024}}

@article{optim_bussi_1,
	author = {Cesari, Andrea and Bottaro, Sandro and Lindorff-Larsen, Kresten and Ban{\'a}s, Pavel and {\v S}poner, Ji{\v r}{\'\i} and Bussi, Giovanni},
	journal = {Journal of chemical theory and computation},
	number = {6},
	pages = {3425--3431},
	publisher = {ACS Publications},
	title = {Fitting corrections to an RNA force field using experimental data},
	volume = {15},
	year = {2019}}

@article{water_tip4p_transfer_dopke,
	author = {D{\"o}pke, Max F and Moultos, Othonas A and Hartkamp, Remco},
	journal = {The Journal of chemical physics},
	number = {2},
	publisher = {AIP Publishing},
	title = {On the transferability of ion parameters to the TIP4P/2005 water model using molecular dynamics simulations},
	volume = {152},
	year = {2020}}

@article{tcpc_pitzer_comp,
	author = {Temoltzi-Avila, Javier and Iglesias-Silva, Gustavo A and Ramos-Estrada, Mariana},
	journal = {Industrial \& Engineering Chemistry Research},
	number = {31},
	pages = {10684--10700},
	publisher = {ACS Publications},
	title = {Comparison among Pitzer model and solvation models. Calculation of osmotic and activity coefficients and dilution enthalpy for single-electrolyte aqueous solutions},
	volume = {57},
	year = {2018}}

@article{tcpc_2,
	author = {Ge, Xinlei and Wang, Xidong and Zhang, Mei and Seetharaman, Seshadri},
	journal = {Journal of Chemical \& Engineering Data},
	number = {2},
	pages = {538--547},
	publisher = {ACS Publications},
	title = {Correlation and prediction of activity and osmotic coefficients of aqueous electrolytes at 298.15 K by the modified TCPC model},
	volume = {52},
	year = {2007}}

@article{desnity_target_pavel,
	author = {Cruces Chamorro, Victor and Jungwirth, Pavel and Martinez-Seara, Hector},
	journal = {The Journal of Physical Chemistry Letters},
	number = {10},
	pages = {2922--2928},
	publisher = {ACS Publications},
	title = {Building water models compatible with charge scaling molecular dynamics},
	volume = {15},
	year = {2024}}

@article{tcpc_1,
	author = {Lin, Cheng-Long and Lee, Liang-Sun and Tseng, Hsieng-Cheng},
	journal = {Fluid phase equilibria},
	number = {1},
	pages = {57--79},
	publisher = {Elsevier},
	title = {Thermodynamic behavior of electrolyte solutions: Part I. Activity coefficients and osmotic coefficients of binary systems},
	volume = {90},
	year = {1993}}

@article{pitzer_eqn_2,
	author = {Krumgalz, BS and Pogorelsky, R and Iosilevskii, Ya A and Weiser, A and Pitzer, Kenneth S},
	journal = {Journal of solution chemistry},
	number = {8},
	pages = {849--875},
	publisher = {Springer},
	title = {Ion interaction approach for volumetric calculations for solutions of single electrolytes at 25 C},
	volume = {23},
	year = {1994}}

@article{pitzer_eqn_1,
	author = {Pitzer, Kenneth S},
	journal = {The Journal of physical chemistry},
	number = {2},
	pages = {268--277},
	publisher = {ACS Publications},
	title = {Thermodynamics of electrolytes. I. Theoretical basis and general equations},
	volume = {77},
	year = {1973}}

@article{vpo_pearce,
	author = {Pearce, JN and Nelson, AF},
	journal = {Journal of the American Chemical Society},
	number = {9},
	pages = {3544--3555},
	publisher = {ACS Publications},
	title = {The vapor pressures of aqueous solutions of lithium nitrate and the activity coefficients of some alkali salts in solutions of high concentration at 25},
	volume = {54},
	year = {1932}}

@article{eutectic_conde,
	author = {Conde, Manuel R},
	journal = {International Journal of Thermal Sciences},
	number = {4},
	pages = {367--382},
	publisher = {Elsevier},
	title = {Properties of aqueous solutions of lithium and calcium chlorides: formulations for use in air conditioning equipment design},
	volume = {43},
	year = {2004}}

@book{apparant_nernst,
	author = {Nernst, W. and Lehfeldt, R.A.},
	lccn = {agr08000935},
	publisher = {Macmillan and Company, limited},
	title = {Theoretical Chemistry from the Standpoint of Avogadro's Rule \& Thermodynamics, chapter VIII},
	url = {https://books.google.it/books?id=kAxDAAAAIAAJ},
	year = {1904}}

@article{apparent_vercher,
	author = {Vercher, Ernesto and Solsona, Solange and V{\'a}zquez, M Isabel and Mart{\'\i}ez-Andreu, Antoni},
	journal = {Fluid phase equilibria},
	number = {1},
	pages = {95--111},
	publisher = {Elsevier},
	title = {Apparent molar volumes of lithium chloride in 1-propanol+ water in the temperature range from 288.15 to 318.15 K},
	volume = {209},
	year = {2003}}

@article{apparant_pitzer,
	author = {Krumgalz, Boris S and Pogorelsky, Rita and Pitzer, Kenneth S},
	journal = {Journal of Physical and Chemical Reference Data},
	number = {2},
	pages = {663--689},
	publisher = {American Institute of Physics for the National Institute of Standards and~{\ldots}},
	title = {Volumetric Properties of Single Aqueous Electrolytes from Zero to Saturation Concentration at 298.15 K Represented by Pitzer's Ion-Interaction Equations},
	volume = {25},
	year = {1996}}

@article{apparant_vaslow,
	author = {Vaslow, Fred},
	journal = {The Journal of Physical Chemistry},
	number = {7},
	pages = {2286--2294},
	publisher = {ACS Publications},
	title = {The apparent molal volumes of the alkali metal chlorides in aqueous solution and evidence for salt-induced structure transitions},
	volume = {70},
	year = {1966}}

@article{apparant_owen,
	author = {Owen, Benton B and Brinkley Jr, Stuart R},
	journal = {Annals of the New York Academy of Sciences},
	number = {4},
	pages = {753--764},
	publisher = {Wiley Online Library},
	title = {Extrapolation of apparent molal properties of strong electrolytes},
	volume = {51},
	year = {1949}}

@article{apparant_masson,
	author = {Masson, D Orme},
	journal = {The London, Edinburgh, and Dublin Philosophical Magazine and Journal of Science},
	number = {49},
	pages = {218--235},
	publisher = {Taylor \& Francis},
	title = {XXVIII. Solute molecular volumes in relation to solvation and ionization},
	volume = {8},
	year = {1929}}

@article{apparant_baxter,
	author = {Baxter, Gregory Paul},
	journal = {Journal of the American Chemical Society},
	number = {6},
	pages = {922--940},
	publisher = {ACS Publications},
	title = {Changes in Volume upon Solution in Water of the Halogen Salts of the Alkalis.},
	volume = {33},
	year = {1911}}

@article{apparant_green,
	author = {Green, W Heber},
	journal = {Journal of the Chemical Society, Transactions},
	pages = {2023--2048},
	publisher = {Royal Society of Chemistry},
	title = {CCII.---Studies on the viscosity and conductivity of some aqueous solutions. Part I. Solutions of sucrose, hydrogen chloride, and lithium chloride},
	volume = {93},
	year = {1908}}

@article{apparant_density_klugman,
	author = {Klugman, I Yu},
	journal = {Russian journal of electrochemistry},
	number = {4},
	pages = {415--420},
	title = {Partial and apparent molar volumes of aqueous solutions of 1: 1 electrolytes},
	volume = {38},
	year = {2002}}

@book{apparant_density_sohnel_novotny,
	author = { S{\"o}hnel, Otakar and Novotn{\'y}, Petr},
	publisher = {Elsevier},
	series = {Journal of Organometallic Chemistry Library},
	title = {Densities of Aqueous Solutions of Inorganic Substances},
	year = {1985}}

@book{apparant_density_wolf,
	author = {Wolf, A.V.},
	lccn = {65002831},
	publisher = {Hoeber Medical Division, Harper \& Row},
	title = {Aqueous Solutions and Body Fluids: Their Concentration Properties and Conversion Tables},
	url = {https://books.google.it/books?id=j_5qAAAAMAAJ},
	year = {1966}}

@techreport{python_net_x,
	author = {Hagberg, Aric and Swart, Pieter J and Schult, Daniel A},
	institution = {Los Alamos National Laboratory (LANL)},
	title = {Exploring network structure, dynamics, and function using NetworkX},
	year = {2007}}

@misc{python_pyvoro,
	author = {Joe Jordan and Andrey Sobolev},
	howpublished = {https://github.com/joe-jordan/pyvoro},
	title = {pyvoro},
	year = {2014}}

@article{python_numpy,
	author = {Charles R. Harris and K. Jarrod Millman and St{\'{e}}fan J. van der Walt and Ralf Gommers and Pauli Virtanen and David Cournapeau and Eric Wieser and Julian Taylor and Sebastian Berg and Nathaniel J. Smith and Robert Kern and Matti Picus and Stephan Hoyer and Marten H. van Kerkwijk and Matthew Brett and Allan Haldane and Jaime Fern{\'{a}}ndez del R{\'{i}}o and Mark Wiebe and Pearu Peterson and Pierre G{\'{e}}rard-Marchant and Kevin Sheppard and Tyler Reddy and Warren Weckesser and Hameer Abbasi and Christoph Gohlke and Travis E. Oliphant},
	doi = {10.1038/s41586-020-2649-2},
	journal = {Nature},
	month = sep,
	number = {7825},
	pages = {357--362},
	publisher = {Springer Science and Business Media {LLC}},
	title = {Array programming with {NumPy}},
	url = {https://doi.org/10.1038/s41586-020-2649-2},
	volume = {585},
	year = {2020}}

@article{python_scipy,
	adsurl = {https://rdcu.be/b08Wh},
	author = {Virtanen, Pauli and Gommers, Ralf and Oliphant, Travis E. and Haberland, Matt and Reddy, Tyler and Cournapeau, David and Burovski, Evgeni and Peterson, Pearu and Weckesser, Warren and Bright, Jonathan and {van der Walt}, St{\'e}fan J. and Brett, Matthew and Wilson, Joshua and Millman, K. Jarrod and Mayorov, Nikolay and Nelson, Andrew R. J. and Jones, Eric and Kern, Robert and Larson, Eric and Carey, C J and Polat, {\.I}lhan and Feng, Yu and Moore, Eric W. and {VanderPlas}, Jake and Laxalde, Denis and Perktold, Josef and Cimrman, Robert and Henriksen, Ian and Quintero, E. A. and Harris, Charles R. and Archibald, Anne M. and Ribeiro, Ant{\^o}nio H. and Pedregosa, Fabian and {van Mulbregt}, Paul and {SciPy 1.0 Contributors}},
	doi = {10.1038/s41592-019-0686-2},
	journal = {Nature Methods},
	pages = {261--272},
	title = {{{SciPy} 1.0: Fundamental Algorithms for Scientific Computing in Python}},
	volume = {17},
	year = {2020}}

@article{software_voro++,
	author = {Rycroft, Chris},
	title = {VORO++: A three-dimensional Voronoi cell library in C++},
	year = {2009}}

@article{bussi_baro,
	author = {Bernetti, Mattia and Bussi, Giovanni},
	journal = {The Journal of Chemical Physics},
	number = {11},
	publisher = {AIP Publishing},
	title = {Pressure control using stochastic cell rescaling},
	volume = {153},
	year = {2020}}

@article{bussi_thermo,
	author = {Bussi, Giovanni and Donadio, Davide and Parrinello, Michele},
	journal = {The Journal of chemical physics},
	number = {1},
	publisher = {AIP Publishing},
	title = {Canonical sampling through velocity rescaling},
	volume = {126},
	year = {2007}}

@article{kb_2,
	author = {Hall, DG},
	journal = {Transactions of the Faraday Society},
	pages = {2516--2524},
	publisher = {Royal Society of Chemistry},
	title = {Kirkwood-Buff theory of solutions. An alternative derivation of part of it and some applications},
	volume = {67},
	year = {1971}}

@article{kb_1,
	author = {Kirkwood, John G and Buff, Frank P},
	journal = {The Journal of chemical physics},
	number = {6},
	pages = {774--777},
	publisher = {American Institute of Physics},
	title = {The statistical mechanical theory of solutions. I},
	volume = {19},
	year = {1951}}

@article{raman_water_2_expt,
	author = {Walrafen, GE and Fisher, MR and Hokmabadi, MS and Yang, W-H},
	journal = {The Journal of chemical physics},
	number = {12},
	pages = {6970--6982},
	publisher = {American Institute of Physics},
	title = {Temperature dependence of the low-and high-frequency Raman scattering from liquid water},
	volume = {85},
	year = {1986}}

@article{raman_water_1_expt,
	author = {Walrafen, GE and Fisher, MR and Hokmabadi, MS and Yang, W-H},
	journal = {The Journal of chemical physics},
	number = {12},
	pages = {6970--6982},
	publisher = {American Institute of Physics},
	title = {Temperature dependence of the low-and high-frequency Raman scattering from liquid water},
	volume = {85},
	year = {1986}}

@article{raman_mcr_geissler,
	author = {Smith, Jared D and Saykally, Richard J and Geissler, Phillip L},
	journal = {Journal of the American Chemical Society},
	number = {45},
	pages = {13847--13856},
	publisher = {ACS Publications},
	title = {The effects of dissolved halide anions on hydrogen bonding in liquid water},
	volume = {129},
	year = {2007}}

@article{raman_mcr_review,
	author = {de Juan, Anna and Tauler, Roma},
	journal = {Analytica Chimica Acta},
	pages = {59--78},
	publisher = {Elsevier},
	title = {Multivariate Curve Resolution: 50 years addressing the mixture analysis problem--A review},
	volume = {1145},
	year = {2021}}

@article{raman_theory_scaling,
	author = {Tikhonov, Denis S and Gordiy, Igor and Iakovlev, Danila A and Gorislav, Alisa A and Kalinin, Mikhail A and Nikolenko, Sergei A and Malaskeevich, Ksenia M and Yureva, Karina and Matsokin, Nikita A and Schnell, Melanie},
	journal = {ChemPhysChem},
	number = {23},
	pages = {e202400547},
	publisher = {Wiley Online Library},
	title = {Harmonic scale factors of fundamental transitions for dispersion-corrected quantum chemical methods},
	volume = {25},
	year = {2024}}

@article{raman_mcr_1_dor,
	author = {Perera, Pradeep and Wyche, Melanie and Loethen, Yvette and Ben-Amotz, Dor},
	journal = {Journal of the American Chemical Society},
	number = {14},
	pages = {4576--4577},
	publisher = {ACS Publications},
	title = {Solute-induced perturbations of solvent-shell molecules observed using multivariate Raman curve resolution},
	volume = {130},
	year = {2008}}

@article{raman_mcr_2_dor,
	author = {Fega, Kathryn Rebecca and Wilcox, Avid Scott and Ben-Amotz, Dor},
	journal = {Applied spectroscopy},
	number = {3},
	pages = {282--288},
	publisher = {SAGE Publications Sage UK: London, England},
	title = {Application of Raman multivariate curve resolution to solvation-shell spectroscopy},
	volume = {66},
	year = {2012}}

@article{raman_red_shift_skinner,
	author = {Bakker, Huib J and Skinner, James L},
	journal = {Chemical reviews},
	number = {3},
	pages = {1498--1517},
	publisher = {ACS Publications},
	title = {Vibrational spectroscopy as a probe of structure and dynamics in liquid water},
	volume = {110},
	year = {2010}}

@article{raman_red_shift_markus,
	author = {Rozenberg, Mark and Loewenschuss, Aharon and Marcus, Yizhak},
	journal = {Physical Chemistry Chemical Physics},
	number = {12},
	pages = {2699--2702},
	publisher = {Royal Society of Chemistry},
	title = {An empirical correlation between stretching vibration redshift and hydrogen bond length},
	volume = {2},
	year = {2000}}

@article{raman_red_shift_buck,
	author = {Buckingham, AD and Del Bene, JE and McDowell, SAC},
	journal = {Chemical Physics Letters},
	number = {1-3},
	pages = {1--10},
	publisher = {Elsevier},
	title = {The hydrogen bond},
	volume = {463},
	year = {2008}}

@article{tess_7,
	author = {Richards, Frederic M},
	journal = {Journal of molecular biology},
	number = {1},
	pages = {1--14},
	publisher = {Elsevier},
	title = {The interpretation of protein structures: total volume, group volume distributions and packing density},
	volume = {82},
	year = {1974}}

@book{graph_paths_main,
	author = {Bondy, John Adrian and Murty, Uppaluri Siva Ramachandra and others},
	publisher = {Macmillan London},
	title = {Graph theory with applications, chapter 1},
	volume = {290},
	year = {1976}}

@article{graph_paths_1,
	author = {Biondi, E and Divieti, L and Guardabassi, G},
	journal = {Canadian Journal of Mathematics},
	number = {1},
	pages = {22--35},
	publisher = {Cambridge University Press},
	title = {Counting paths, circuits, chains, and cycles in graphs: a unified approach},
	volume = {22},
	year = {1970}}

@article{graph_ring_goet,
	author = {Goetzke, K and Klein, H-J},
	journal = {Journal of non-crystalline solids},
	number = {2},
	pages = {215--220},
	publisher = {Elsevier},
	title = {Properties and efficient algorithmic determination of different classes of rings in finite and infinite polyhedral networks},
	volume = {127},
	year = {1991}}

@article{graph_ring_prim,
	author = {Yuan, Xianglong and Cormack, AN},
	journal = {Computational materials science},
	number = {3},
	pages = {343--360},
	publisher = {Elsevier},
	title = {Efficient algorithm for primitive ring statistics in topological networks},
	volume = {24},
	year = {2002}}

@article{graph_roux,
	author = {Han, Kyungreem and Venable, Richard M and Bryant, Anne-Marie and Legacy, Christopher J and Shen, Rong and Li, Hui and Roux, Beno{\^\i}t and Gericke, Arne and Pastor, Richard W},
	journal = {The Journal of Physical Chemistry B},
	number = {4},
	pages = {1484--1494},
	publisher = {ACS Publications},
	title = {Graph--theoretic analysis of monomethyl phosphate clustering in ionic solutions},
	volume = {122},
	year = {2018}}

@article{graph_choi_3,
	author = {Choi, Jun-Ho and Choi, Hyung Ran and Jeon, Jonggu and Cho, Minhaeng},
	journal = {The Journal of chemical physics},
	number = {15},
	publisher = {AIP Publishing},
	title = {Ion aggregation in high salt solutions. VII. The effect of cations on the structures of ion aggregates and water hydrogen-bonding network},
	volume = {147},
	year = {2017}}

@article{graph_choi_2,
	author = {Choi, Jun-Ho and Cho, Minhaeng},
	journal = {The Journal of chemical physics},
	number = {10},
	publisher = {AIP Publishing},
	title = {Ion aggregation in high salt solutions. IV. Graph-theoretical analyses of ion aggregate structure and water hydrogen bonding network},
	volume = {143},
	year = {2015}}

@article{graph_choi_1,
	author = {Choi, Jun-Ho and Cho, Minhaeng},
	journal = {The Journal of chemical physics},
	number = {15},
	publisher = {AIP Publishing},
	title = {Ion aggregation in high salt solutions. II. Spectral graph analysis of water hydrogen-bonding network and ion aggregate structures},
	volume = {141},
	year = {2014}}

@book{intro_hansen_mcdonald,
	author = {Hansen, Jean-Pierre and McDonald, Ian Ranald},
	publisher = {Academic press},
	title = {Theory of simple liquids: with applications to soft matter, chapter 3},
	year = {2013}}

@article{software_vmd_main,
	author = {Humphrey, William and Dalke, Andrew and Schulten, Klaus},
	journal = {Journal of molecular graphics},
	number = {1},
	pages = {33--38},
	publisher = {Elsevier},
	title = {VMD: visual molecular dynamics},
	volume = {14},
	year = {1996}}

@article{orca_d4,
	author = {Caldeweyher, Eike and Mewes, Jan-Michael and Ehlert, Sebastian and Grimme, Stefan},
	journal = {Physical Chemistry Chemical Physics},
	number = {16},
	pages = {8499--8512},
	publisher = {Royal Society of Chemistry},
	title = {Extension and evaluation of the D4 London-dispersion model for periodic systems},
	volume = {22},
	year = {2020}}

@article{orca_vdw,
	author = {Caldeweyher, Eike and Ehlert, Sebastian and Hansen, Andreas and Neugebauer, Hagen and Spicher, Sebastian and Bannwarth, Christoph and Grimme, Stefan},
	journal = {The Journal of chemical physics},
	number = {15},
	publisher = {AIP Publishing},
	title = {A generally applicable atomic-charge dependent London dispersion correction},
	volume = {150},
	year = {2019}}

@article{orca_libxc,
	author = {Lehtola, Susi and Steigemann, Conrad and Oliveira, Micael JT and Marques, Miguel AL},
	journal = {SoftwareX},
	pages = {1--5},
	publisher = {Elsevier},
	title = {Recent developments in libxc---A comprehensive library of functionals for density functional theory},
	volume = {7},
	year = {2018}}

@article{orca_d3_grimme,
	author = {Caldeweyher, Eike and Bannwarth, Christoph and Grimme, Stefan},
	journal = {The Journal of chemical physics},
	number = {3},
	publisher = {AIP Publishing},
	title = {Extension of the D3 dispersion coefficient model},
	volume = {147},
	year = {2017}}

@article{orca_hessian,
	author = {Bykov, Dmytro and Petrenko, Taras and Izs{\'a}k, R{\'o}bert and Kossmann, Simone and Becker, Ute and Valeev, Edward and Neese, Frank},
	journal = {Molecular Physics},
	number = {13-14},
	pages = {1961--1977},
	publisher = {Taylor \& Francis},
	title = {Efficient implementation of the analytic second derivatives of Hartree--Fock and hybrid DFT energies: a detailed analysis of different approximations},
	volume = {113},
	year = {2015}}

@article{orca_ri,
	author = {Neese, Frank},
	journal = {Journal of computational chemistry},
	number = {14},
	pages = {1740--1747},
	publisher = {Wiley Online Library},
	title = {An improvement of the resolution of the identity approximation for the formation of the Coulomb matrix},
	volume = {24},
	year = {2003}}

@article{voro_paper,
	author = {Voronoi, Georges},
	journal = {Journal f{\"u}r die reine und angewandte Mathematik},
	number = {136},
	pages = {67--182},
	publisher = {De Gruyter Berlin, New York},
	title = {Nouvelles applications des param{\`e}tres continus {\`a} th{\'e}orie des formes quadratiques. Deuxi{\`e}me M{\'e}moire. Recherches sur les parall{\'e}lo{\`e}dres primitifs.},
	volume = {1909},
	year = {1909}}

@article{voro_laguerre_main,
	author = {Aurenhammer, Franz},
	journal = {SIAM journal on computing},
	number = {1},
	pages = {78--96},
	publisher = {SIAM},
	title = {Power diagrams: properties, algorithms and applications},
	volume = {16},
	year = {1987}}

@article{tess_6,
	author = {Voloshin, Vladimir P and Kim, Alexandra V and Medvedev, Nikolai N and Winter, Roland and Geiger, Alfons},
	journal = {Biophysical Chemistry},
	pages = {1--9},
	publisher = {Elsevier},
	title = {Calculation of the volumetric characteristics of biomacromolecules in solution by the Voronoi--Delaunay technique},
	volume = {192},
	year = {2014}}

@article{tess_5,
	author = {Voloshin, Vladimir P and Medvedev, Nikolai N and Andrews, Maximilian N and Burri, R Reddy and Winter, Roland and Geiger, Alfons},
	journal = {The Journal of Physical Chemistry B},
	number = {48},
	pages = {14217--14228},
	publisher = {ACS Publications},
	title = {Volumetric properties of hydrated peptides: Voronoi--Delaunay analysis of molecular simulation runs},
	volume = {115},
	year = {2011}}

@article{tess_3,
	author = {Ashbaugh, Henry S and Barnett, J Wesley and Saltzman, Alexander and Langrehr, Mae and Houser, Hayden},
	journal = {The Journal of Physical Chemistry B},
	number = {13},
	pages = {3242--3250},
	publisher = {ACS Publications},
	title = {Connections between the anomalous volumetric properties of alcohols in aqueous solution and the volume of hydrophobic association},
	volume = {122},
	year = {2017}}

@article{tess_2,
	author = {Yang, Yongjian and Tokunaga, Hirofumi and Ono, Madoka and Hayashi, Kazutaka and Mauro, John C},
	journal = {Scripta Materialia},
	pages = {1--5},
	publisher = {Elsevier},
	title = {Understanding the molar volume of alkali-alkaline earth-silicate glasses via Voronoi polyhedra analysis},
	volume = {166},
	year = {2019}}

@article{tess_1,
	author = {Kadtsyn, Evgenii D and Nichiporenko, Vladislav A and Medvedev, Nikolai N},
	journal = {Journal of Molecular Liquids},
	pages = {118173},
	publisher = {Elsevier},
	title = {Volumetric properties of solutions on the perspective of Voronoi tessellation},
	volume = {349},
	year = {2022}}

@article{orca_shark,
	author = {Neese, Frank},
	journal = {Journal of Computational Chemistry},
	number = {3},
	pages = {381--396},
	publisher = {Wiley Online Library},
	title = {The SHARK integral generation and digestion system},
	volume = {44},
	year = {2023}}

@article{raman_theory_hess,
	author = {Neugebauer, Johannes and Reiher, Markus and Kind, Carsten and Hess, Bernd A},
	journal = {Journal of computational chemistry},
	number = {9},
	pages = {895--910},
	publisher = {Wiley Online Library},
	title = {Quantum chemical calculation of vibrational spectra of large molecules---Raman and IR spectra for Buckminsterfullerene},
	volume = {23},
	year = {2002}}

@article{software_orca,
	author = {Neese, Frank},
	journal = {Wiley Interdisciplinary Reviews: Computational Molecular Science},
	number = {2},
	pages = {e70019},
	publisher = {Wiley Online Library},
	title = {Software update: the ORCA program system---version 6.0},
	volume = {15},
	year = {2025}}

@article{software_gmx,
	author = {Abraham, Mark James and Murtola, Teemu and Schulz, Roland and P{\'a}ll, Szil{\'a}rd and Smith, Jeremy C and Hess, Berk and Lindahl, Erik},
	journal = {SoftwareX},
	pages = {19--25},
	publisher = {Elsevier},
	title = {GROMACS: High performance molecular simulations through multi-level parallelism from laptops to supercomputers},
	volume = {1},
	year = {2015}}

@article{intro_review_vandervegt,
	author = {Van Der Vegt, Nico FA and Haldrup, Kristoffer and Roke, Sylvie and Zheng, Junrong and Lund, Mikael and Bakker, Huib J},
	journal = {Chemical reviews},
	number = {13},
	pages = {7626--7641},
	publisher = {ACS Publications},
	title = {Water-mediated ion pairing: Occurrence and relevance},
	volume = {116},
	year = {2016}}

@article{intro_review_marcus,
	author = {Marcus, Yizhak},
	journal = {Chemical reviews},
	number = {3},
	pages = {1346--1370},
	publisher = {ACS Publications},
	title = {Effect of ions on the structure of water: structure making and breaking},
	volume = {109},
	year = {2009}}

@article{mac_mayer_2,
	author = {Vafaei, Shaghayegh and Tomberli, Bruno and Gray, CG},
	journal = {The Journal of chemical physics},
	number = {15},
	publisher = {AIP Publishing},
	title = {McMillan-Mayer theory of solutions revisited: Simplifications and extensions},
	volume = {141},
	year = {2014}}

@article{mac_mayer_1,
	author = {McMillan Jr, William G and Mayer, Joseph E},
	journal = {The Journal of Chemical Physics},
	number = {7},
	pages = {276--305},
	publisher = {American Institute of Physics},
	title = {The statistical thermodynamics of multicomponent systems},
	volume = {13},
	year = {1945}}

@article{ion_jc_2008,
	author = {Joung, In Suk and Cheatham III, Thomas E},
	journal = {The journal of physical chemistry B},
	number = {30},
	pages = {9020--9041},
	publisher = {ACS Publications},
	title = {Determination of alkali and halide monovalent ion parameters for use in explicitly solvated biomolecular simulations},
	volume = {112},
	year = {2008}}

@article{ion_madrid_2019,
	author = {Zeron, IM and Abascal, JLF and Vega, C},
	journal = {The Journal of chemical physics},
	number = {13},
	publisher = {AIP Publishing},
	title = {A force field of Li+, Na+, K+, Mg2+, Ca2+, Cl-, and SO42- in aqueous solution based on the TIP4P/2005 water model and scaled charges for the ions},
	volume = {151},
	year = {2019}}

@article{density_target_madrid_nacl,
	author = {Benavides, AL and Portillo, MA and Chamorro, VC and Espinosa, JR and Abascal, JLF and Vega, C},
	journal = {The Journal of chemical physics},
	number = {10},
	publisher = {AIP Publishing},
	title = {A potential model for sodium chloride solutions based on the TIP4P/2005 water model},
	volume = {147},
	year = {2017}}

@article{water_tip4p_05_abascal,
	author = {Abascal, Jose LF and Vega, Carlos},
	journal = {The Journal of chemical physics},
	number = {23},
	publisher = {AIP Publishing},
	title = {A general purpose model for the condensed phases of water: TIP4P/2005},
	volume = {123},
	year = {2005}}

@article{density_target_opls_aa,
	author = {Jorgensen, William L and Maxwell, David S and Tirado-Rives, Julian},
	journal = {Journal of the american chemical society},
	number = {45},
	pages = {11225--11236},
	publisher = {ACS Publications},
	title = {Development and testing of the OPLS all-atom force field on conformational energetics and properties of organic liquids},
	volume = {118},
	year = {1996}}

@article{density_target_booth,
	author = {Boothroyd, Simon and Madin, Owen C and Mobley, David L and Wang, Lee-Ping and Chodera, John D and Shirts, Michael R},
	journal = {Journal of chemical theory and computation},
	number = {6},
	pages = {3577--3592},
	publisher = {ACS Publications},
	title = {Improving force field accuracy by training against condensed-phase mixture properties},
	volume = {18},
	year = {2022}}

@article{water_jorg_tip3_4p,
	author = {Jorgensen, William L and Chandrasekhar, Jayaraman and Madura, Jeffry D and Impey, Roger W and Klein, Michael L},
	journal = {The Journal of chemical physics},
	number = {2},
	pages = {926--935},
	publisher = {American Institute of Physics},
	title = {Comparison of simple potential functions for simulating liquid water},
	volume = {79},
	year = {1983}}

@article{water_spce,
	author = {Berendsen, Herman JC and Grigera, J-Ra{\'u}l and Straatsma, Tjerk P},
	journal = {Journal of physical chemistry},
	number = {24},
	pages = {6269--6271},
	publisher = {ACS Publications},
	title = {The missing term in effective pair potentials},
	volume = {91},
	year = {1987}}

@article{density_target_gaff,
	author = {Wang, Junmei and Wolf, Romain M and Caldwell, James W and Kollman, Peter A and Case, David A},
	journal = {Journal of computational chemistry},
	number = {9},
	pages = {1157--1174},
	publisher = {Wiley Online Library},
	title = {Development and testing of a general amber force field},
	volume = {25},
	year = {2004}}

@article{density_target_parm94,
	author = {Cornell, Wendy D and Cieplak, Piotr and Bayly, Christopher I and Gould, Ian R and Merz, Kenneth M and Ferguson, David M and Spellmeyer, David C and Fox, Thomas and Caldwell, James W and Kollman, Peter A},
	journal = {Journal of the American Chemical Society},
	number = {19},
	pages = {5179--5197},
	publisher = {ACS Publications},
	title = {A second generation force field for the simulation of proteins, nucleic acids, and organic molecules},
	volume = {117},
	year = {1995}}

@article{density_target_alkane,
	author = {Ungerer, Philippe and Beauvais, Christele and Delhommelle, J{\'e}r{\^o}me and Boutin, Anne and Rousseau, Bernard and Fuchs, Alain H},
	journal = {The Journal of Chemical Physics},
	number = {12},
	pages = {5499--5510},
	publisher = {American Institute of Physics},
	title = {Optimization of the anisotropic united atoms intermolecular potential for n-alkanes},
	volume = {112},
	year = {2000}}

@article{density_target_opls,
	author = {Jorgensen, William L and Madura, Jeffry D and Swenson, Carol J},
	journal = {Journal of the American Chemical Society},
	number = {22},
	pages = {6638--6646},
	publisher = {ACS Publications},
	title = {Optimized intermolecular potential functions for liquid hydrocarbons},
	volume = {106},
	year = {1984}}

@article{density_target_roux,
	author = {Boulanger, Eliot and Huang, Lei and Rupakheti, Chetan and MacKerell Jr, Alexander D and Roux, Beno{\^\i}t},
	journal = {Journal of chemical theory and computation},
	number = {6},
	pages = {3121--3131},
	publisher = {ACS Publications},
	title = {Optimized Lennard-Jones parameters for druglike small molecules},
	volume = {14},
	year = {2018}}

@article{hof,
	author = {Hofmeister, Franz},
	journal = {Archiv f{\"u}r experimentelle Pathologie und Pharmakologie},
	number = {1},
	pages = {1--30},
	publisher = {Springer},
	title = {Zur lehre von der wirkung der salze: Dritte mittheilung},
	volume = {25},
	year = {1888}}

@article{intro_voigt,
  title={Chemistry of salts in aqueous solutions: Applications, experiments, and theory},
  author={Voigt, Wolfgang},
  journal={Pure and Applied Chemistry},
  volume={83},
  number={5},
  pages={1015--1030},
  year={2011},
  publisher={De Gruyter}
}

\end{document}